\documentclass[fleqn,usenatbib]{mnras}

\usepackage{newtxtext,newtxmath}

\usepackage[T1]{fontenc}

\DeclareRobustCommand{\VAN}[3]{#2}
\let\VANthebibliography\thebibliography
\def\thebibliography{\DeclareRobustCommand{\VAN}[3]{##3}\VANthebibliography}

\usepackage{graphicx}
\usepackage{amsmath}
\usepackage{orcidlink}
\usepackage{cancel}

\newcommand{\arepo}{\textsc{arepo}}
\newcommand{\fable}{\textsc{fable}}
\newcommand{\sE}{\textsc{fable}-sE}
\newcommand{\vl}{\texttt{VoroLite++}}
\newcommand{\msun}{\mathrm{M}_{\odot}}
\newcommand{\rvir}{R_{\mathrm{vir}}}
\newcommand{\mvir}{M_{\mathrm{vir}}}
\newcommand{\mbh}{M_{\mathrm{BH}}}
\newcommand{\fedd}{f_{\mathrm{Edd}}}
\newcommand{\hi}{\ion{H}{i}}

\title[Outflows in a super-Eddington quasar at $z \gtrsim 6$]{Outflows in a super-Eddington quasar drive clumpy circumgalactic medium and extended H$\alpha$ nebulae at $z \gtrsim 6$}
\author[L. Tortora et al.]{
Lucas Tortora$^{\orcidlink{0009-0005-8040-8325}}$,$^{1}$\thanks{Corresponding author: lt589@cam.ac.uk (LT)}
Tiago Costa$^{\orcidlink{0000-0002-6748-2900}}$,$^{2}$
Debora Sijacki$^{\orcidlink{0000-0002-3459-0438}}$$^{1}$
and Jake S. Bennett$^{\orcidlink{0000-0002-8573-2993}}$$^{3}$
\\
$^{1}$Institute of Astronomy and Kavli Institute for Cosmology, Cambridge, University of Cambridge, Madingley Road, Cambridge CB3 0HA, UK\\
$^{2}$School of Mathematics, Statistics and Physics, Newcastle University, Newcastle upon Tyne, NE1 7RU, UK\\
$^{3}$School of Physics \& Astronomy, University of Nottingham, University Park, Nottingham, NG7 2RD, UK
}

\date{MNRAS, accepted}

\pubyear{2026}

\begin{document}
\label{firstpage}
\pagerange{\pageref{firstpage}--\pageref{lastpage}}
\maketitle

\begin{abstract}
The discovery of gargantuan black holes with masses exceeding a billion solar masses at $z\gtrsim6$ suggests rapid black hole growth and significant energy input into their host galaxies in the early Universe. With \textit{JWST} probing previously unseen phases of the interstellar (ISM) and circumgalactic (CGM) medium around $z > 6$ quasars, detailed theoretical studies can now be directly confronted with observations.
We use zoom-in simulations of a massive protocluster at $z\sim6$, employing both the fiducial \fable{} galaxy formation model and modifications that allow earlier black hole seeding and mildly super-Eddington accretion.
The central quasar remains Compton-thick throughout most of its evolution, with the obscuration arising from the ISM of its compact host galaxy. 
The onset of sufficiently strong quasar feedback drives a `blow-out' episode, clearing out escape channels for ionizing radiation and leaving the central engine unobscured.
This leads to a complete transformation of the CGM, whereby powerful, metal-enriched outflows produce a population of cold, fast, neutral clumps, significantly increasing the covering fraction of neutral hydrogen in the host halo.
Radiative transfer calculations performed with a new ray-tracing code show that the CGM responds to quasar activity through the formation of H$\alpha$ nebulae, whose size and luminosity increase with the strength of quasar feedback and decrease with obscuration level. Enhanced early black hole growth thus fundamentally reshapes the ISM and CGM of $z\sim6$ quasars, leaving clear observable signatures in their obscuration, neutral hydrogen distribution, and extended H$\alpha$ emission.
\end{abstract}




\section{Introduction}

The existence of supermassive black holes (SMBHs) is well-established, as they are ubiquitously found in the centre of local galaxies \citep[see, e.g.,][]{genzel+_2000, ghez+_2000, EHT+_m87_2019, EHT+_sgra*_2022, GRAVITY+_sgra*_2022}. Over the last decades, observations have accumulated strong evidence linking the properties of SMBHs with several properties of their galactic hosts.
These are reflected in scaling relations between, e.g., black hole and stellar masses, and velocity dispersion \citep[e.g.,][]{magorrian+_1998, kormendy_ho_2013, McConnell_Ma_2013, Reines_Volonteri_2015, Greene_Strader_Ho_2020}. These connections may suggest co-evolution between SMBHs and the galaxies in which they reside, mediated by a variety of physical processes such as gas accretion, galaxy mergers, and feedback \citep[e.g.,][]{Haehnelt_Natarajan_Rees_1998, Silk_Rees_1998, King_2003, dimatteo_springel_hernquist_2005, sijacki+_2007, hopkins+_2008, Fabian_2012, king_pounds_2015, Somerville_Davé_2015, Alexander+_2025}.

The observed population of active galactic nuclei (AGN) are thought to be powered by accretion of gas onto SMBHs, with quasars (or QSOs) corresponding to the brightest and most extreme manifestation of AGN \citep[e.g.,][]{Schmidt_1963, Salpeter_1964, Lynden-Bell_1969, Rees_1984, Inayoshi-rev_2020, wang+_2021, Fan_Banados_Simcoe_2023}.
Their exceptional luminosities ($L_{\mathrm{bol}}\gtrsim10^{46}$--$10^{48}~{\rm erg~s^{-1}}$) make them detectable out to very early cosmic times. Over 300 quasars have been discovered at redshifts $z\gtrsim6$ across the entire electromagnetic spectrum with a variety of instruments \citep[see][for comprehensive reviews]{Inayoshi-rev_2020, Fan_Banados_Simcoe_2023, Banados_2026}. The masses of the accreting SMBHs powering such quasars are usually inferred by converting the widths of broad emission lines in their spectra into masses using single-epoch virial relations \citep[e.g.,][]{peterson+_2004, vestergaard_peterson_2006, vestergaard_osmer_2009}. At $z>6$, the \ion{C}{iv} and \ion{Mg}{ii} lines have been extensively used in the past \citep[e.g.,][]{Schindler+_2020, farina+_2022, mazzucchelli+_xqr30_2023}, while the advent of the \textit{James Webb Space Telescope} \citep[\textit{JWST};][]{JWST_2006, JWST_2023} has enabled the use of the more reliable H$\alpha$ and H$\beta$ lines \citep[e.g.,][]{kokorev+_2023, marshall+_2023, yang_j+_2023, Maiolino+_2024_JADES, juodzbalis+_2025_JADES}. These observations confirm the implication that $z>6$ quasars are powered by SMBHs with masses in excess of $\gtrsim10^9$\,M$_{\sun}$.

The very existence of this population of likely extremely massive SMBHs is surprising, as it is non-trivial to explain how such objects could assemble within the first billion years of the Universe \citep[e.g.,][]{Turner_1991, Haiman_Loeb_2001, Inayoshi-rev_2020}. Massive black hole seeds ($M_{\mathrm{BH, \, seed}}\approx 10^4$--$10^6~{\rm M_{\sun}}$) may provide a `head-start' \citep[see, e.g.,][]{sijacki_springel_haehnelt_2009, Volonteri_Habouzit_Colpi_2021}, though the subsequent growth of the black hole has to proceed at close to the Eddington rate for a significant fraction of a Hubble time \citep[e.g.,][]{wang+_2021, Maiolino+_2024_GNz11} or, alternatively, through multiple bursts of super-Eddington accretion \citep[e.g.,][]{Volonteri_Rees_2005, Madau_Haardt_Dotti_2014, Volonteri_Silk_Dubus_2015, Inayoshi-rev_2020, bennett+_2024}. Even if black holes are seeded massive, the rapid assembly of high-redshift SMBHs requires sustained, high accretion rates despite feedback and/or gas supply limitations to be able to explain the detections of luminous $z\gtrsim6$ quasars.

Puzzlingly, recent measurements indicate that the luminous lifetimes of $z > 6$ quasars are often remarkably short \citep[e.g.,][]{eilers+_2024}. Hints of short quasar lifetimes are provided by quasar proximity zones, which constrain total luminous timescales through the response of the surrounding intergalactic medium (IGM) to its net absorption of ionizing photons \citep[][]{eilers_hennawi_davies_2018, eilers+_2020, eilers+_2021, davies+_2020, morey+_2021, durovcikova+_2024, Durovcikova+_2026_BEES}. This observational data may support a picture in which a large majority of the accretion onto $z\gtrsim6$ SMBHs is primarily obscured or inherently radiatively inefficient \citep[e.g.,][]{hopkins+_2005, RamosAlmeida_Ricci_2017, Hickox_Alexander_2018, Vito+_2018, Circosta+_2019, davies_hennawi_eilers_2019, D'Amato+_2020, gilli+_2022, satyavolu+_2023a_obs, yang_g+_2023, Bulichi+_2026, Leung+_2026}. Cosmological, hydrodynamical simulations of galaxy formation and black hole growth indeed show that high ($\sim80-90$~per cent) obscuration fractions are achieved in the special environments of high-redshift quasars, providing further support for obscured SMBH growth in the early Universe \citep[e.g.,][]{costa+_2018b_quenching_sf, Trebitsch+_2019, ni+_2020, bennett+_2024}.

Theoretical models also predict that accretion releases enormous amounts of energy. The resulting radiative and mechanical feedback should power large-scale outflows that disrupt the interstellar (ISM) and circumgalactic (CGM) media of haloes hosting $z\gtrsim6$ quasars \citep[e.g.,][]{sijacki_springel_haehnelt_2009, Fabian_2012, costa+_2014, costa_sijacki_haehnelt_2015, costa+_2018b_quenching_sf, costa_pakmor_springel_2020, costa+_2022, king_pounds_2015, bieri+_2017, barai+_2018, smidt+2018, zhu+_2022, ward+_2024}, while simultaneously setting the local scaling relations as a result of the co-evolution between SMBHs and their host galaxies \citep[e.g.,][]{Haehnelt_Natarajan_Rees_1998, Silk_Rees_1998, sijacki+_2007, volonteri_reines_2016, fiore+_2017, valentini_gallerani_ferrara_2021}. While detections of broad absorption-line features suggest that small-scale, vigorous ($v\sim0.1c$) winds capable of launching galactic outflows exist in the nuclei of quasar host galaxies \citep[e.g.,][]{meyer_bosman_ellis_2019, choi+_2020, Schindler+_2020, wang+_2021, yang_j+_2021, bischetti+_2022, bischetti+_2023}, observational evidence for the latter has proven controversial. Historically, \citet[][]{Maiolino+_2012} and \citet[][]{cicone+_2015} were the first to report on an extended ($R\sim15~{\rm kpc}$) outflow traced by [\ion{C}{ii}] 158\,$\mu\mathrm{m}$ emission in a $z=6.4$ quasar, with further evidence for outflows corroborated by stacked analyses of high-redshift QSOs \citep[][]{bischetti+_2019, stanley+_2019}. Other studies including similar targets, however, refuted these results and found no strong evidence for outflows \citep[][]{decarli+_2018, novak+_2020}. Most recently, several studies leveraging new Atacama Large Millimetre Array (ALMA) data as well as the capabilities of \textit{JWST} have revealed that outflows are ubiquitous around high-redshift quasars, and may be more extreme than seen in their counterparts in the local Universe \citep[e.g.,][]{marshall+_2023, yang_j+_2023, bischetti+_2025, spilker+_2025, Liu+_2026}.

Another source of ambiguity is the environment of $z \gtrsim 6$ quasars. In the concordance $\Lambda$CDM cosmology, the brightest high-redshift quasars should reside in the most massive haloes that lie within overdensities of the cosmic density field, effectively tracing the first large-scale structure in the Universe \citep[e.g.,][]{efstathiou_rees_1988, volonteri_rees_2006, latif_ferrara_2016, Inayoshi-rev_2020}. These regions rapidly assemble into dark matter haloes with exceptionally high masses ($M_{\mathrm{halo}}\gtrsim10^{12}\,\msun$ at $z\approx6$), where both frequent mergers and sustained, `smooth' cosmic gas inflows facilitate the growth of the first SMBHs by accretion \citep[e.g.,][]{li+_2007, sijacki_springel_haehnelt_2009, dimatteo+_2012, costa+_2014}. In fact, this is practically a requirement for simulations to grow SMBHs on par with the most massive objects observed in our Universe \citep[e.g.,][]{bennett+_2024, costa_2024}. A direct observational prediction of this hypothesis is the presence of an anomalously large quantity of companion galaxies in the vicinity of high-redshift quasars \citep[][]{munoz_loeb_2008, tinker+_2010}. Despite sustained and targeted efforts to search for these companions in the supposedly overdense environments of luminous quasars, no clear consensus has been reached. Mixed and contradictory results are reported in the literature, wherein some quasar fields appear overdense while others are not, and may even be underdense \citep[e.g.,][]{banados+_2013, Mazzucchelli+_2017, meyer+_2022, kashino+_2023, wang+_2023, eilers+_2024}.

A better understanding of quasar environments, lifetimes, and feedback processes could potentially be attained by probing the CGM, which forms the primary reservoir for the gas flows that regulate SMBH growth \citep[see][for reviews]{tumlinson_peebles_werk_2017, f-g_oh_2023}.
As a result, a variety of observational signatures spanning the electromagnetic spectrum, accessible with both current and upcoming facilities, could be used to constrain the physical properties of quasars and their gaseous haloes at high redshift \citep[see, e.g.,][and references therein]{costa+_2022, bennett+_2024}.
For example, extended X-ray absorption and emission, and Sunyaev-Zeldovich effect \citep[SZ;][]{Sunyaev_Zeldovich_1970, Sunyaev_Zeldovich_1972, Sunyaev_Zeldovich_1980} maps can help constrain the strength of AGN feedback \citep[e.g.,][]{Athena+_2013, brownson+_2019, DiMascolo+2023, bogdan+_2023}. The CGM of high-redshift quasars is also now being targeted by \textit{JWST} in the rest-optical and rest-NIR, supplemented by new data from ALMA and the Multi Unit Spectroscopic Explorer (MUSE) at the VLT, to offer new probes of the ISM, dust, outflows and their interplay with the extended gaseous halo \citep[e.g.,][]{ding+_2023, galbiati+_2023, tripodi+_2023, bischetti+_2025, Liu+_2025, Valentino+_2026, Durovcikova+_2026_BEES, Wolf+_2026}. Additionally, \textit{Euclid} \citep[][]{Euclid_2022} is accelerating the discovery of high-redshift quasars, which will be crucial to probe a greater variety of objects and environments \citep[e.g.,][]{euclid_barnett+_2019, Belladitta+_2026_Euclid, Yang_D+_2026_Euclid}.

In the last decade, multiple models have been used to simulate black hole formation, accretion and feedback in a fully cosmological context, largely matching local constraints despite varying numerical implementations \citep[e.g.,][]{habouzit+_2021, habouzit+_2022_low_z, feldmann_bieri_rev}. At high redshift, however, these models disagree and predict fundamentally different physical pictures \citep[][]{habouzit+_2022_high_z}. To better resolve black hole feeding and its impact on quasar hosts, the `zoom-in' technique is often used, targeting the rarest, most massive overdensities found in large dark-matter-only simulations and resimulating them at higher resolution. While these works have successfully produced multiple SMBHs with $\mbh\gtrsim10^9\,\msun$ at $z\gtrsim6$ \citep[e.g.,][]{sijacki_springel_haehnelt_2009, costa+_2014, lupi+_2019, valentini_gallerani_ferrara_2021, bhowmick+_2022, bhowmick+_2026, zhu+_2022, bennett+_2024, husko+_2025, quadri+_2025}, it remains challenging for numerical simulations to form the most extreme SMBHs exceeding $10^{10}\,\msun$ at $z\sim6$, as estimated for the record-holder J010013.02+280225.8 \citep[][]{wu+_2015}. Where achieved, simulations disagree on the exact formation mechanisms: in some cases, this can only be achieved by turning AGN feedback off \citep[][]{valentini_gallerani_ferrara_2021}, while in others, very steep requirements for the halo ($M_{\mathrm{halo}}>10^{13}\,\msun$) and seed ($M_{\mathrm{BH, \, seed}}=10^6\,\msun$) masses must be invoked \citep[][]{zhu+_2022}.

Theoretical predictions for the growth mechanisms and impact on the gaseous halo of ultramassive SMBHs at high redshift thus remain scarce. In \citet[][]{bennett+_2024}, small modifications to the \fable{} galaxy formation model \citep[][]{fable_henden+_2018, fable_bigwood+_2025}, such as earlier seeding of black holes and mildly super-Eddington accretion, enabled the formation and growth of a gargantuan $\mbh=1.3\times10^{10}\,\msun$ object at $z=6$, whose progenitors are also consistent with the brightest quasars at $z>7$ \citep[e.g.,][]{Larson+_2023, Maiolino+_2024_GNz11}. It was noted that feedback from the super-Eddington accreting SMBH was much stronger, launching powerful metal-enriched outflows that significantly perturbed the CGM of the host halo. However, the impact of the quasar radiation field was not explicitly studied. The aim of this paper is to provide an in-depth study and characterisation of the outflows launched by super-Eddington accreting SMBHs and the response of the CGM of the host haloes by modelling the radiation from quasar phases in post-processing. This regime is now widely-discussed, as super-Eddington accretion and the related outflows may explain recent \textit{JWST} findings \citep[][]{Hu+_2022, madau_haardt_2024, Pacucci_Narayan_2024, Tortosa+_2024, Saccheo+_2025, Suh+_2025, inayoshi_kimura_noda_2025, Inayoshi_Maiolino_2025, Chaikin+_2026, Madau_2026}. By connecting these scales to the obscuration stage of the central engine, we aim to gain a better understanding of the relation between black hole feeding and phenomena on scales of the host halo, and to provide new constraints for interpreting current and upcoming observations of the CGM of $z \gtrsim 6$ quasars.

This paper is organized in the following manner. In Section~\ref{sec:methods}, we introduce the simulations used in this work and present our methodology for performing the radiative transfer simulations with a novel ray-tracing code. In Section~\ref{sec:results}, we present our findings, focusing firstly on characterising the outflows found in the simulations before showing the outcome of the radiative transfer calculations. We discuss our results in the context of other works in Section~\ref{sec:discussions}, and finally summarise our work and present our conclusions in Section~\ref{sec:conclusions}.

\section{Methodology}\label{sec:methods}

\subsection{Simulations}

In this paper, we analyse the suite of three zoom-in simulations introduced in \citet{bennett+_2024}. The simulations are performed with the cosmological, hydrodynamical code \arepo{} \citep[][]{arepo_springel_2010, arepo_pakmor+_2016, arepo_weinberger+_2020}. The code follows gas hydrodynamics on a moving mesh constructed from the Voronoi tessellation of a discrete number of mesh-generating points. To identify bound structures in the simulations, we use an on-the-fly friends-of-friends \citep[][]{Davis_1985} and \textsc{subfind} \citep[][]{springel+_2001, dolag+_2009} algorithms. We briefly review the setup of the zoom-in simulations here and refer the reader to the previous paper for additional details. 

The zoom-in simulations target the largest halo from the Millennium cosmological box at redshift $z=6$ \citep[][]{springel+_2005}, with varying models of baryonic physics. At $z=6$, its virial mass\footnote{We use the spherical overdensity criterion to compute virial quantities, such that $\mvir$ refers to the mass contained within $\rvir$, the radius within which the average density is 200 times the critical density of the universe at a given redshift.} is $\mvir\approx7\times10^{12} \, \msun$ and virial radius is $\rvir\approx85~{\rm proper~kpc}$. While the $z \, = \, 0$ descendants of such haloes can range from galaxy groups to rich clusters \citep{Angulo_2012}, the halo targeted by our simulations evolves into a rich galaxy cluster with $\mvir = 4.6\times 10^{15}\,\msun$ by $z=0$. The mass resolution of the gas and stellar component is $1.45\times10^6\,h^{-1}\,\msun$, and $6.62\times10^6\,h^{-1}\,\msun$ for dark matter. 
The comoving softening lengths are $2.5~{\rm kpc}$ and adaptive, such that the mass-weighted mean radii of cells at a distance of $2~{\rm proper~kpc}$ from the halo centre are $\approx 0.2~{\rm proper~kpc}$, and they increase to $\approx 1~{\rm proper~kpc}$ at $50~{\rm proper~kpc}$ from the centre. We assume cosmological parameters are consistent with constraints set by \emph{WMAP} \citep{wmap_spergel+_2007}, namely: $\Omega_{\Lambda}=0.75$, $\Omega_{\mathrm{m}}=0.25$, $\Omega_{\mathrm{b}}=0.045$, $\sigma_8=0.9$ and $h=0.73$. Throughout the rest of the paper, units are given in proper (physical) coordinates. For further details on the setup of the zoom-in simulations, see \citet{sijacki_springel_haehnelt_2009, bennett+_2024}.

The simulations analysed here use varying flavours of the \fable{} galaxy formation model \citep[][]{fable_henden+_2018, fable_bigwood+_2025}, which itself builds upon the Illustris model \citep[][]{vogelsberger+_2013, illustris_vogelsberger+_2014, illustris_genel+_2014, illustris_torrey+_2014, illustris_sijacki+_2015}. The models for gas cooling, star formation and stellar feedback are identical to those in \fable{} for all three simulations.

We use a similar nomenclature to \citet{bennett+_2024}, wherein the three simulations in our suite are labelled `NoAGN', `\fable{}' and `\sE{}' (formerly `Reference'). In the NoAGN run, all physics related to black holes (as introduced in the next paragraphs) is excluded, while in the \fable{} and \sE{} runs, respectively, the fiducial and modified black hole physics are used \citep[see][for details]{bennett+_2024}.

In the fiducial model, black holes are seeded with a mass $M_{\mathrm{BH, \ seed}}=10^5\,h^{-1}\,\msun$ at the position of the gravitational potential minimum of every dark matter halo once its mass exceeds $\mvir>5\times10^{10}\,h^{-1}\,\msun$. Black hole particles are modelled as collisionless sink particles that can grow through mergers with other black holes and gas accretion. The latter is parametrized by an Eddington-limited Bondi-Hoyle-Lyttleton-like model \citep[][]{bondi_hoyle_1944, bondi_1952} wherein
\begin{equation}\label{eq:BHL}
    \dot{M}_{\mathrm{BH}}=\alpha \frac{4\pi \, G^2 \, \mbh^2 \, \rho}{c_{s}^3}\,,
\end{equation}
where $\dot{M}_{\mathrm{BH}}$ is the black hole accretion rate, $G$ is the gravitational constant, $\mbh$ is the mass of the black hole, $\rho$ and $c_s$ are, respectively, the density and sound speed of the surrounding gas (evaluated from kernel-averaging over the 32 nearest gas cells in the simulation), and $\alpha$ is a numerical boost factor. The latter is introduced to artificially boost the accretion of gas onto black holes, which in reality should reside in a denser, multi-phase environment that cannot be captured when the ISM is not explicitly resolved \citep[see e.g.][for discussions]{springel_dimatteo_hernquist_2005}; \fable{} employs a boost factor $\alpha=100$.

The luminosity of the AGN is defined as $L_{\rm bol}=\epsilon_{\rm r}\,\dot{M}_{\rm BH} \, c^2$ where $\epsilon_{\mathrm{r}}$ is the radiative efficiency. In \fable{}, this value is set to $\epsilon_{\mathrm{r}}=0.1$ \citep[][]{Soltan_1982, Rees_1984, Yu_Tremaine_2002, Shankar+_2004, fable_henden+_2018}. The fiducial model assumes the accretion rate is capped at the Eddington limit due to this radiation output, which (assuming ionized hydrogen) is given by
\begin{equation}\label{eq:edd-limit}
    \dot{M}_{\mathrm{Edd}}=\frac{4\pi \, G\, \mbh \, m_{\mathrm{p}}}{\epsilon_\mathrm{r} \, \sigma_{\mathrm{T}}\, c}\,,
\end{equation}
with $m_{\mathrm{p}}$ the proton mass, $c$ the speed of light and $\sigma_{\mathrm{T}}$ the Thomson scattering cross-section. We define the Eddington ratio as $\fedd \, = \, \dot{M}_{\mathrm{BH}}\,/\, \dot{M}_{\mathrm{Edd}}$.
Finally, an additional `pressure criterion' modulates the accretion rate, such that $\dot{M}_{\mathrm{BH}}$ is reduced when large ($\gtrsim10^9~\msun$) SMBHs are embedded in low-density background gas to avoid unphysical inflation of hot, diffuse bubbles \citep{vogelsberger+_2013}.

Feedback from AGN is treated via a dual-mode model depending on the normalised black hole accretion rate $\fedd$. At $\fedd>0.01$, the luminous quasar mode operates, wherein a fraction $\epsilon_{\mathrm{f}}$ of the luminosity $L_{\rm bol}$ powering the bright quasar is isotropically injected as thermal energy in the surroundings of the black hole. The energy injection rate is thus
\begin{equation}\label{eq:fb-quasar}
    \dot{E}_{\mathrm{FB, \ quasar \ mode}} = \epsilon_{\mathrm{f}} \, L_{\mathrm{bol}} = \epsilon_{\mathrm{f}} \, \epsilon_{\mathrm{r}}\,\dot{M}_{\mathrm{BH}} \, c^2\,.
\end{equation}

To mitigate artificial cooling losses in the neighbouring gas, a duty cycle on the quasar mode is invoked, such that the feedback energy is stored for $25$~Myr and then injected into the gas in a single feedback event \citep[see][for further details]{fable_henden+_2018}. At $\fedd<0.01$, the AGN operates in the (radiatively inefficient) radio mode, in which hot bubbles are injected into the gas at a distance of $50 \, h^{-1}~{\rm kpc}$ from the black hole as if inflated by AGN jets, following \citet{sijacki+_2007}. Bubbles are periodically injected when the gain in mass exceeds $\delta_{\mathrm{BH}}=\delta \mbh \, /\, \mbh$ with mechanical heating efficiency $\epsilon_{\mathrm{m}}$, such that
\begin{equation}\label{eq:radio}
    \dot{E}_{\mathrm{FB, \ radio \ mode}} = \epsilon_{\mathrm{m}} \, \epsilon_{\mathrm{r}}\, \delta M_{\mathrm{BH}} \, c^2\,.
\end{equation}
Fiducially, we use the following values in the \fable{} run: $\epsilon_{\rm f}=0.1, \epsilon_{\rm m}=0.8$ and $\delta_{\rm BH}=0.01$.

Due to insufficient numerical resolution, it is very challenging for cosmological simulations to resolve dynamical friction correctly. To avoid numerical wandering of black holes, they are kept at the centre of their host halo by repositioning them at the gravitational potential minimum of particles within their smoothing length at each active time-step \citep[][]{springel_dimatteo_hernquist_2005}. Furthermore, black hole mergers are assumed to happen instantaneously when two black holes are within each other's smoothing lengths. These effects likely combine to produce overly efficient and overabundant mergers in these models \citep[see, e.g.,][for a detailed discussion]{buttigieg+_2025}, though we note that a majority of the final black hole mass comes from accretion.

The \sE{} simulation introduces several modifications to the physics of black holes from the fiducial \fable{} model to promote early black hole growth. We refer the reader to \citet[][section 2.3]{bennett+_2024} for details. We emphasize that these simulations were not designed to `correctly' grow extremely massive SMBHs, but rather to offer a ``plausible'' pathway to their assembly and to study the impact that such objects would have on their surroundings.

In \sE{}, black hole seeds are placed in smaller mass haloes with $\mvir>10^{9}\,h^{-1}\,\msun$, and we allow for mildly super-Eddington accretion of gas, with $\fedd=2$ up from the Eddington-limited case of \fable{} (with $\fedd=1$). Additionally, we slightly reduce the numerical feedback coupling efficiency in \sE{} to $\epsilon_{\rm f}=0.05$, down from $\epsilon_{\rm f}=0.1$ used in \fable{}, and note that $\epsilon_{\rm f}$ is kept fixed independently of Eddington ratio. This parameter choice encourages earlier accretion and pushes feedback-regulated growth to higher black hole masses \citep[][]{bennett+_2024}. Importantly, we note that lowering $\epsilon_{\rm f}$ does not reduce the total injected feedback energy in the simulations; in fact, we show later that \textit{stronger} feedback is produced despite the lower coupling. In combination, these modifications produce an SMBH with the gargantuan mass of $\mbh=1.3\times10^{10}\,\msun$ at $z=6$ for the \sE{} run, one order of magnitude higher than the already large $\mbh=1.4\times10^{9}\,\msun$ of the \fable{} run. Detailed discussions on the black hole mass and its evolution in these simulations, including the effect of the host assembly history, are presented in \citet[][]{bennett+_2024}.

\subsection{Tracing neutral hydrogen}

The purpose of the present study is to characterize, in post-processing, the ionization state of hydrogen when exposed to radiation from a quasar. Gas in and around the quasar host halo is also irradiated by stars and background sources, which we do not explicitly model in the radiative transfer simulations. In this work, we thus make the approximation that gas is in ionization equilibrium with a homogeneous, redshift-dependent ionizing ultraviolet background (UVB) from \citet[][]{puchwein+_2019}. From the output of the hydrodynamical simulations, we compute the predicted ionization state of hydrogen and use this as the initial conditions for the radiative transfer simulations. We present a detailed summary of this approach in Appendix~\ref{app:neutral-hydrogen}.

One important subtlety of this procedure concerns the star-forming gas, whose temperature is defined by the imposed pressure of the `effective' polytropic equation of state \citep[][]{springel_hernquist_2003} and is not representative of a physical temperature. For this reason, we manually impose $T_{\rm ISM}=10^4~{\rm K}$ for star-forming gas throughout this paper, including for the recombination and collisional ionization states used in the radiative transfer and to compute emission properties \citep[see, e.g.,][]{rahmati+_2013}.

We make use of the covering fraction to quantify the distribution of neutral hydrogen around the simulated halo. We mainly consider strong \hi{} absorbers, specifically Damped Lyman-$\alpha$ systems (DLAs) with $N_{\hi{}}>10^{20.3}$\,cm$^{-2}$. The cumulative \hi{} covering fraction is defined as
\begin{equation}\label{eq:fcov}
    f_{\mathrm{cov}}(<R)=\frac{A_{\mathrm{abs}}(<R)}{\pi\,R^2} \, ,
\end{equation}
where $A_{\mathrm{abs}}$ is the area covered by lines of sight above the DLA threshold within the field of view defined by $R$. This is equivalent to the probability of finding DLAs within a circular aperture of radius $R$ \citep[see][]{f-g+_2015, f-g+_2016, Rahmati+_2015, tortora+_2024}.

From the ionization state of the gas, we compute the H$\alpha$ luminosity and surface brightness to produce mock images. For every gas cell in the simulations, we compute the H$\alpha$ emissivity given as
\begin{equation}\label{eq:halpha}
    \varepsilon_{\rm H\alpha} = h\nu_{\rm H\alpha} \, \alpha^{\rm eff}_{\rm H\alpha}(T) \, n_{\rm e} \, n_{\ion{H}{ii}} \ {\rm erg \, s^{-1} \, cm^{-3}} \, ,
\end{equation}
where $h\nu_{\rm H\alpha}$ is the energy of an H$\alpha$ photon (6564.6 \AA), $\alpha^{\rm eff}_{\rm H\alpha}(T)$ is the effective Case B recombination emissivity coefficient for the H$\alpha$ line \citep[][]{Pequignot_Petitjean_Boisson_1991}, and $n_{\rm e}$ and $n_{\ion{H}{ii}}$ are the electron and ionized hydrogen number densities, respectively. To obtain the H$\alpha$ luminosity, we simply project the emissivity along the line of sight; for the surface brightness, we also account for cosmological dimming ($\propto(1+z)^{-4}$) and the pixel size in arcsec$^2$.

\subsection{Radiative transfer in post-processing}\label{sec:vorolite}

We post-process the simulations with the new radiative transfer tool \vl{} (Costa et al. in prep.) designed specifically to perform ray-tracing on a Voronoi grid.
Rays are cast from a single source into \arepo{} cells within a chosen spherical volume centred on the source. 
The number of rays is set such that every cell is guaranteed to be traversed by at least one ray.
Each ray is directed from the source towards the centre of a target cell, defining a unique propagation direction through the mesh. 
The rays are then grouped into equal-area angular bins using a HEALPix tessellation \citep[][]{HEALPix_2005}, allowing a weight to be assigned to each ray such that the total emitted luminosity is conserved and distributed isotropically.

As rays propagate through the Voronoi mesh, neighbour-to-neighbour traversal is performed by calculating the distance along the ray direction to the interfaces separating each adjacent \arepo{} cell. The interface corresponding to the smallest positive distance determines the next cell crossed by the ray, following the strategy presented in \citet[][]{Camps_2013}. The traversed distance and corresponding column density are accumulated cell-by-cell until the ray exits the spherical domain.
Note that since the gas distribution is fixed during post-processing, the geometric ray paths are computed only once at the beginning of the calculation and subsequently reused throughout the radiative transfer step.

The resulting ray trajectories are then used to perform the radiative transfer calculation. Column densities are accumulated cell-by-cell along each ray and subsequently used to evaluate optical depths and compute the attenuation of the photon flux. 
Since the ionization state, and therefore the optical depth, can evolve significantly during a time step, evaluating photon attenuation using the ionization fractions from the beginning of the time step can lead to substantial operator-splitting errors. To prevent this effect, the radiative transfer calculation is first performed using the ionization fractions from the previous time-step, after which a predicted time-averaged ionization state is computed, as in \citet{Mellema_2006}, and subsequently used in a second radiative transfer update prior to the final photo-chemistry step.

The converged photon absorption rates are used to evolve the ionization state of hydrogen. The hydrogen ionization fraction, $x_{\ion{H}{ii}}$, evolves according to 
\begin{equation}\label{eq:hydrogen-ionization-fraction}
\frac{dx_{\ion{H}{ii}}}{dt} \, = \, \Gamma \left(1-x_{\ion{H}{ii}}\right) + C_{\ion{H}{i}} n_{\rm e} \left(1-x_{\ion{H}{ii}}\right) - \alpha_{\ion{H}{ii}} n_{\rm e} x_{\ion{H}{ii}} \, ,
\end{equation}
where $\Gamma$ is the photoionization rate, $C_{\ion{H}{i}}$ is the collisional ionization coefficient, $\alpha_{\ion{H}{ii}}$ is the Case B recombination coefficient, and $n_e$ is the electron density. The recombination and collisional ionization coefficients are calculated using the fitting formulae presented in \citet{Hui_Gnedin_1997}.

Integrating this equation explicitly requires extremely short time-steps. We follow the approach outlined in \citet{Mellema_2006}, whereby the solver computes the equilibrium ionization fraction and corresponding equilibrium timescale in each cell. Under the assumption that the ionization and recombination rates remain constant during a time-step, the ionization equation admits an analytic solution describing exponential relaxation toward the local equilibrium ionization state over the equilibrium timescale. The ionization state is therefore updated analytically over each time-step by relaxing the previous solution towards the equilibrium state. This strategy improves numerical stability for stiff ionization problems and avoids the restrictive time-step requirements associated with explicit integration schemes.

We adopt an infinite-speed-of-light approximation within the illuminated region \citep[see, e.g.,][]{Cantalupo_2011}. A causal propagation condition is additionally imposed by suppressing photon absorption in cells whose light-travel time from the source exceeds the elapsed simulation time, following \citet{Abel_Wandelt_2002}. This approximation is expected to remain accurate once ionization fronts evolve at velocities well below the speed of light.

Photo-heated gas is expected to respond hydrodynamically \citep[e.g.,][]{Ledos_2026}, expanding and thereby modifying its density structure. Since the radiative transfer is performed in post-processing, however, the gas density and temperature remain fixed while evolving the ionization state and radiation field. We therefore neglect photo-heating and assume the radiation source emits an ionizing flux $Q_{\hi{}}$ above the hydrogen ionization threshold ($E_{\rm ion, \, \hi{}}=h \nu_{\rm ion, \, \hi{}}=13.6~{\rm eV}$). The approach adopted here is therefore primarily intended to identify and characterize the spatial distribution of ionized gas associated with \ion{H}{ii} regions. We focus here on relative trends and environmental dependencies rather than precise quantitative predictions for nebular emission.

\begin{figure*}
	\includegraphics[width=\linewidth]{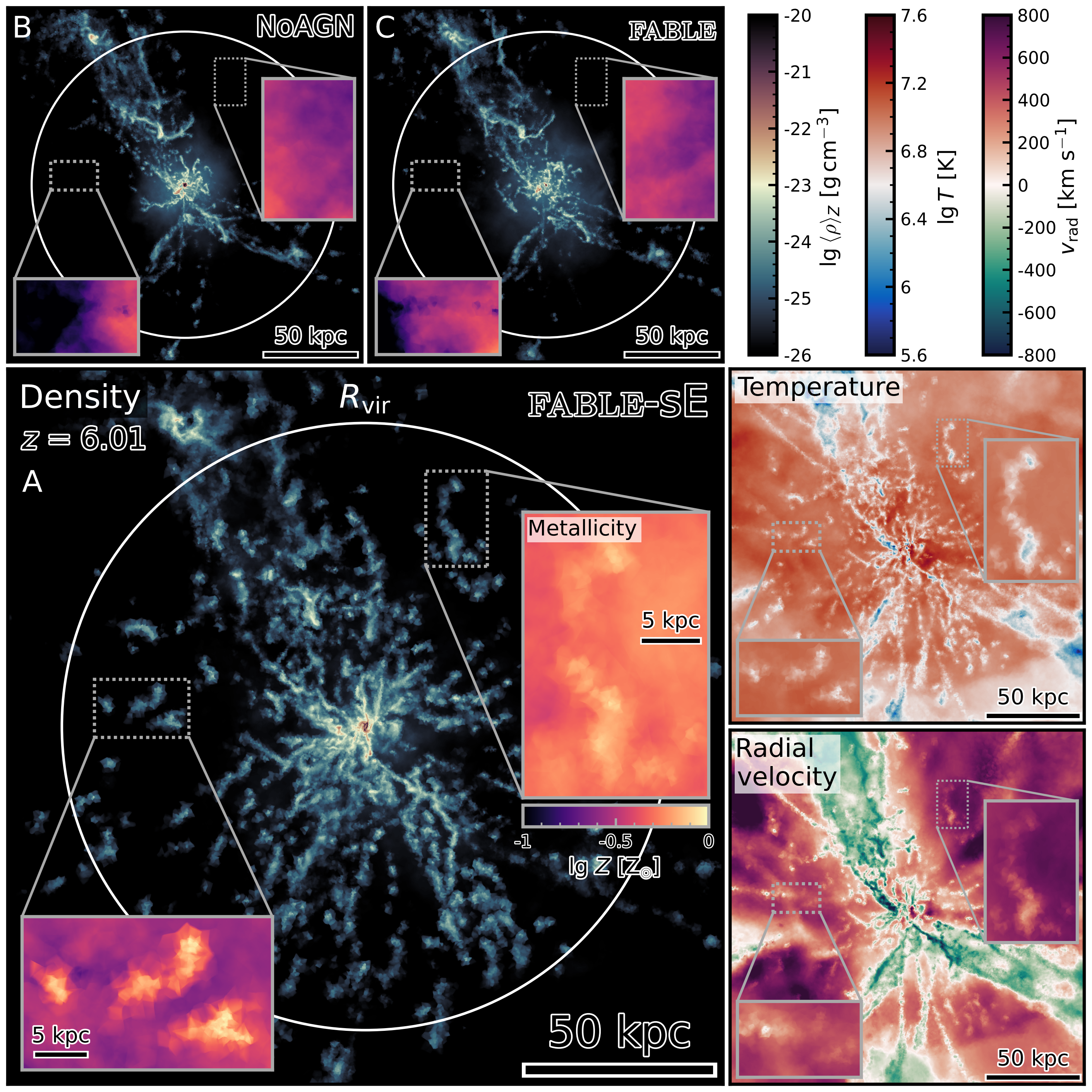}
    \caption{Overview of the suite of simulations used in this work. From the bottom left, clockwise: panels \textbf{A} (\sE{}), \textbf{B} (NoAGN) and \textbf{C} (\fable{}) show the metallicity-weighted gas density of the central halo hosting the quasar in each simulation at redshift $z=6$. The field-of-view in every panel covers $200~{\rm kpc}~\times~200~{\rm kpc}$ with a projection width of $200~{\rm kpc}$, and the white circle shows the virial radius $\rvir \approx 80~{\rm kpc}$. The insets focus on specific regions of the CGM of each halo, where we also display the density-weighted projected metallicity. Panels \textbf{D} and \textbf{E} display, respectively, the density-weighted projected temperature and radial velocity in the \sE{} run, with insets focusing on the same regions highlighted in panel \textbf{A}. \emph{The CGM of the \sE{} halo is profoundly affected and changed by the feedback from the quasar it hosts. In particular, the gas distribution is much clumpier, as shown by the insets highlighting that dense, metal-enriched, cool and outflowing gas is found in the outer ($\lesssim \rvir$) parts of the halo, while absent in the NoAGN and \fable{} runs.}}
    \label{fig:visual_overview}
\end{figure*}

\subsection{Modelling of AGN emission}

To simulate a bright quasar phase, we place an ionizing source at the position of the most massive black hole in the simulations at a given snapshot and explicitly track the ionization state of every cell as photons propagate outward. As outlined in Section~\ref{sec:vorolite}, the radiative transfer is performed in post-processing, such that the \arepo{} mesh and properties of gas cells (e.g., density and temperature) remain fixed. To compute the ionizing flux of the source, we convert the bolometric luminosity predicted from the hydrodynamic simulations $L_{\rm bol}$ into an ionizing photon rate for neutral hydrogen $Q_{\hi{}}$, assuming some form for the intrinsic spectral energy distribution (SED) of the source:
\begin{equation}\label{eq:ioni-photon-rate}
    Q_{\hi{}} = \int_{\nu_{\rm ion, \, \hi{}}}^{\infty} \frac{L_{\nu}}{h \nu} \, d\nu \, ,
\end{equation}
where $L_{\nu} \, d\nu$ is the spectral luminosity density of the source in the frequency range $\nu$ and $\nu+d\nu$ (the SED, such that $L_{\rm bol}=\int_0^\infty L_{\nu}\,d\nu$) and $\nu_{\rm ion, \, \hi{}}$ is the hydrogen ionization threshold. Generally, the SED for AGN is a complex function spanning the entire electromagnetic spectrum, whose shape is a direct reflection of the multi-scale, multi-temperature nature of accreting SMBHs. It thus also depends on the properties of the black holes, such as mass and accretion rate. 

In this work, we adopt a simplified approach wherein we associate an SED with the central SMBH in the simulations from standard models found in the literature. There exist several such models that can be both physically and empirically motivated \citep[e.g.,][]{Collinson+_2015, Castello-Mor_Netzer_Kaspi_2016, Hickox_Alexander_2018, Shen+_2020, Temple_Hewett_Banerji_2021, Su+_2026}. Throughout the paper, we use the physically motivated standard \textsc{qsosed}\footnote{Note that we do not consider the super-Eddington extension \textsc{agnslim} \citep[][]{kubota_done_2019}, as the latter only produces marked changes for $f_{\rm Edd, \, crit}=2.39$, which is above the maximal accretion rate in our simulations.} model \citep[][]{kubota_done_2018}, and provide further details of this methodology in Appendix~\ref{app:ionizing-photon-rate}.

The ionization cross-section $\sigma_{\hi{}}$ used for the radiative transfer is computed as the luminosity-weighted average over the H-ionizing range $[\nu_{\rm ion, \, \hi{}}-\infty]$ \citep[e.g.,][]{Rosdahl+_2013, bieri+_2017}:
\begin{equation}
    \langle\sigma_{\rm HI}\rangle = \frac{\int_{\nu_{\rm ion, \, \hi{}}}^\infty \sigma_{\rm HI}(\nu) \, L_{\nu} \, / \, h\nu \ d\nu}{\int_{\nu_{\rm ion, \, \hi{}}}^\infty L_{\nu} \, / \, h\nu \ d\nu } \, ,
\end{equation}
where $\sigma_{\rm HI}(\nu)$ is the frequency-dependent \hi{} photoionization cross-section from \citet[][]{Verner+_1996}. In principle, this expression should be evaluated for every SED. However, we find that varying $\sigma_{\hi{}}$ does not significantly alter our results, and we thus choose for consistency and simplicity to use the median value for the range of spectra adopted in our work, which is $\langle \sigma_{\hi{}} \rangle=2.126\times10^{-18}~{\rm cm^2}$.

\begin{figure*}
	\includegraphics[width=\linewidth]{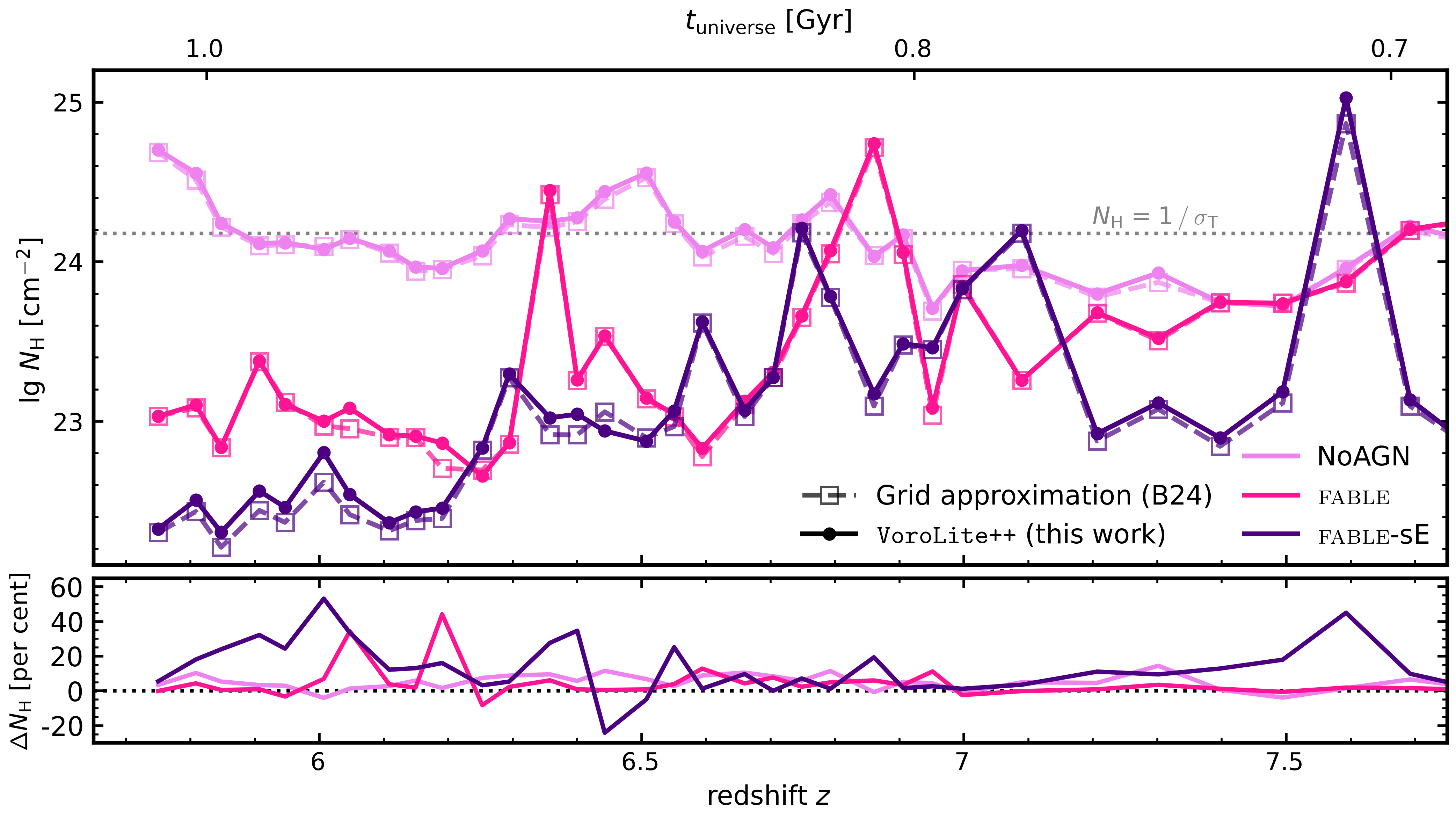}
    \caption{\textit{Upper panel}: Median hydrogen column density ($N_{\rm H}$) along various sightlines covering a sphere of radius $5\,\mathrm{kpc}$ centred on the black hole for redshifts $z\in[5.75-7.8]$ for the simulations used in this work. Dashed lines with square markers highlight results obtained from a simplified approximation adopted in \citealt{bennett+_2024} (B24), see main text for details. Solid lines with point markers show results from tracing through the Voronoi tessellation using \vl{}. The grey dashed vertical line marks the threshold for the gas to be Compton thick ($N_{\mathrm{H}}=1\, / \,\sigma_{\mathrm{T}}$). \textit{Lower panel}: Relative difference between the two methods, defined such that positive values indicate higher columns for \vl{} (in per cent). \emph{Feedback from AGN plays the dominant role in regulating the obscuration level in the centre of the halo, as repeated cycles of `blow-out' episodes clear out gas and reduce central column densities by $\gtrsim2~{\rm dex}$ at $z=5.75$ compared to a run without black hole physics. Column densities obtained by accurately ray-tracing through the \arepo{} mesh with \vl{} are generally higher than those obtained from a simple approximation.}}
    \label{fig:hydrogen_column_density}
\end{figure*}

\subsection{Attenuation by dust}

The radiative transfer simulations performed in this work do not explicitly account for the effects of dust. The predicted $\rm H\alpha$ surface brightness maps are computed directly from the output of the simulations and correspond to the intrinsic (unattenuated) emission. To mimic the effects of dust attenuation and obtain attenuated surface brightness maps, we make the simplified approximation wherein the intrinsic emission is attenuated by a dust screen placed in front of the maps. Specifically, from the intrinsic flux $F_{\rm int}$ we obtain the attenuated flux:
\begin{equation}
    F_{\rm att}=F_{\rm int} \, e^{-\tau_{\rm H\alpha}} \, ,
\end{equation}
where $\tau_{\rm H\alpha}=\kappa_{\rm H\alpha} \Sigma_{\rm dust}$ with $\kappa_{\rm H\alpha}$ the absorption cross section for $\rm H\alpha$ and $\Sigma_{\rm dust}$ the projected dust column density. We use\footnote{See the table provided at: \url{https://www.astro.princeton.edu/~draine/dust/extcurvs/kext_albedo_WD_MW_3.1_60_D03.all}} $\kappa_{\rm H\alpha}=8925~{\rm cm^2~g^{-1}}$ \citep[][]{Draine_2003}. To estimate the dust mass, we make the approximation that dust traces metal-rich and cool gas, specifically that dust makes up 15 per cent of the metal mass of cold ($T<5\times10^4~{\rm K}$ + star-forming) gas \citep[as in][]{DiMascia+_2021, Vito+_2022, bennett+_2024}. This represents a pessimistic estimate for the level of attenuation, such that the results model a worst-case scenario for our approximations, and we leave a more thorough analysis of the effects of dust for future work.

\begin{figure*}
	\includegraphics[width=\linewidth]{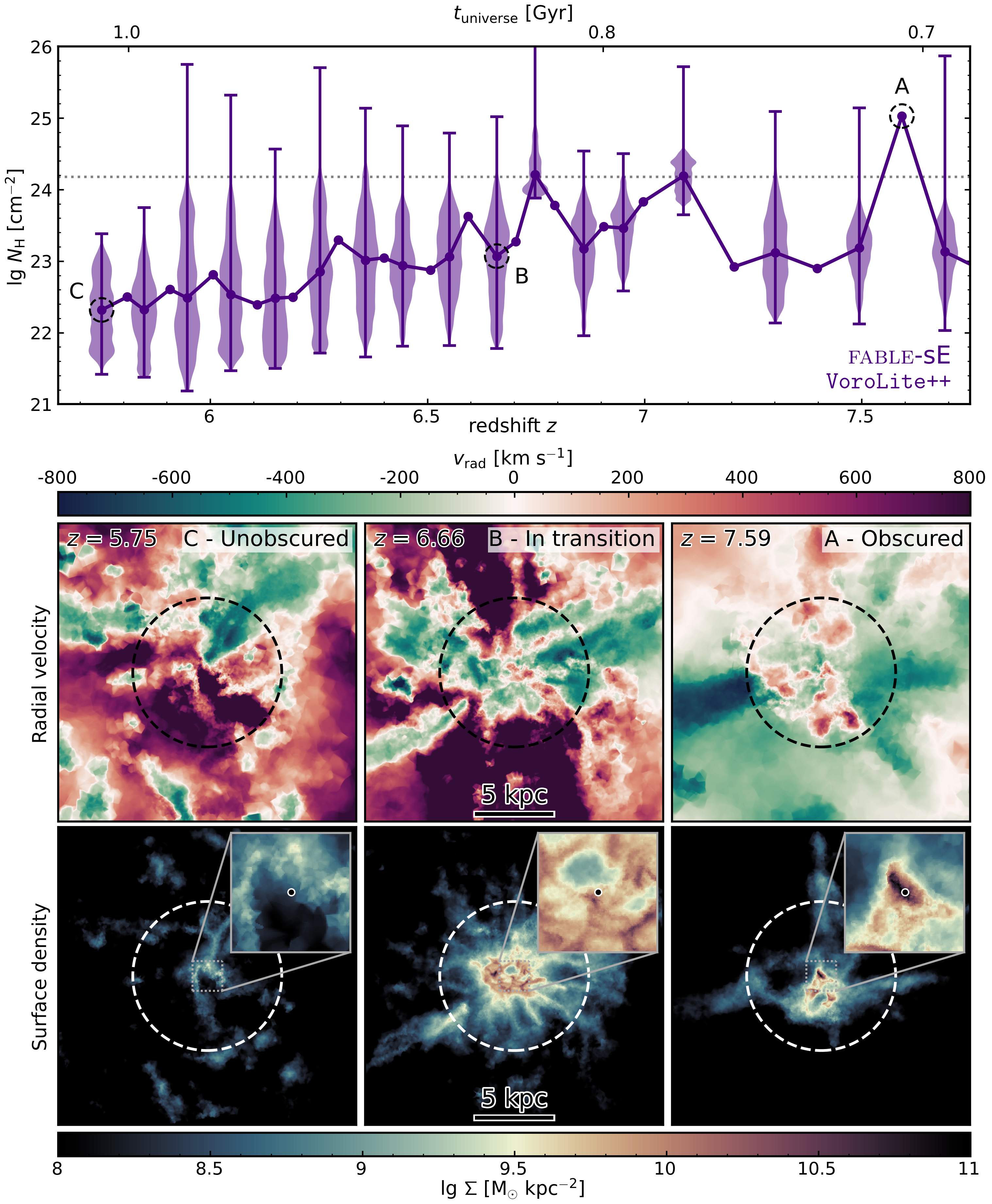}
    \caption{\textit{Upper panel}: Hydrogen column density around the central black hole for the \sE{} run obtained with \vl{} (as in Fig.~\ref{fig:hydrogen_column_density}). The line and associated point markers indicate the median column density along $20000$ sightlines; we also show the histograms highlighting the full distribution of columns, with minimum and maximum values shown with caps. The grey dashed vertical line marks the threshold for the gas to be Compton thick ($N_{\mathrm{H}}=1\, / \,\sigma_{\mathrm{T}}$). 
    \textit{Lower panels}: Density-weighted projected radial velocity and surface density maps at redshifts corresponding to varying black hole obscuration stage, as indexed by the letters `A', `B' and `C' in the upper panel. The dashed circles have a radius of $5~{\rm kpc}$ centred on the SMBH. \emph{Repeated cycles of small blow-out events driven by AGN feedback in \sE{} clear out gas from the centre, leaving the engine completely unobscured by $z=5.75$.}}
    \label{fig:obscuration_detail}
\end{figure*}

\section{Results}\label{sec:results}

\subsection{Overview}

We present a visual inspection of the suite of simulations analysed in this paper in Fig.~\ref{fig:visual_overview}. This includes the density, temperature, radial velocity and metallicity of gas found in a box within $\pm100\,\mathrm{kpc}$ of the halo centre.

The quantities presented are density-weighted projection maps. To obtain the map for a variable $X$ (e.g., temperature, radial velocity, pressure, metallicity, etc), we perform a projection by tracing through the mesh and incorporating the gradient along mesh cells; for every pixel on the final grid, we compute
\begin{equation}\label{eq:vortrace}
    \langle X\rangle= \frac{\int\ X \, \rho \ dl}{\int \rho \ dl} \, ,
\end{equation}
where the integration is done by collecting cell contributions along the chosen line of sight. Column density maps, instead, are obtained by directly projecting the density, e.g., $\Sigma=\int\rho \, dl$.
For panels \textbf{A}, \textbf{B} and \textbf{C}, we display the metallicity-weighted projected density, which is obtained via the following method:
\begin{equation}\label{eq:rho_mz}
    \langle \rho \rangle_Z = \frac{\int \rho^2\,Z \ dl}{\int\rho \, Z \ dl}\, .
\end{equation}
We choose to represent the density in this way to more clearly highlight the difference in the CGM of the \sE{} halo with those of the NoAGN and \fable{} runs.

All simulations show a prominent filamentary structure that connects the central halo to the IGM and cosmic web. These filaments are cold and the gas within them flows towards the centre, fuelling star formation in the galaxy and feeding the SMBH at its core. The rest of the diffuse halo makes up a hot atmosphere of gas, as supernova and AGN feedback eject gas which thermalizes within the host's CGM.
On the smallest scales, we see that the NoAGN simulation produces a very dense, structured, and disc-like central galaxy, whereas the simulations with black hole physics have a disturbed core showing prominent cavities as a result of AGN feedback. On the other hand, although there is a massive black hole in the centre of the \fable{} halo, its CGM visually appears very similar to the one found in the NoAGN run, though there are a number of quantitative differences which we highlight in the next section. This is to be contrasted with the CGM in the \sE{} run, where clear structural differences arise from the very strong feedback from its central quasar. Significant outflows completely alter the halo atmosphere, which is generally hotter with faster gas ($\gtrsim1000~{\rm km \, s^{-1}}$) than in the other runs \citep[see also][]{bennett+_2024}. The filaments are visibly perturbed and fragmented, and the outer ($\lesssim\rvir$) regions of the halo are filled with clumps of dense gas.

These clumps are also found to have markedly different physical properties from the rest of the halo. As highlighted in the insets in panels \textbf{A}, \textbf{D} and \textbf{E} from Fig.~\ref{fig:visual_overview}, clumps are metal-enriched, preferentially outflowing, and are cooler than their surroundings. The presence of cool and dense gas clumps in the outer halo has clear consequences on observational properties of the \sE{} run, which we discuss in Section~\ref{sec:ioni-state}.

\subsection{AGN obscuration with \vl{}}\label{sec:obscuration}

In this section, we zoom into the central parts of the halo to study the obscuration of the black hole.

We compute the total hydrogen column density ($N_{\rm H}$) in two different ways. For the first method, we construct a grid of $\sim12000$ equally-spaced rays on the unit sphere using HEALPix \citep[][with \texttt{NSIDE=32}]{HEALPix_2005}, further subdivided into 1024 logarithmically-spaced radial bins, from the centre out to $d_{\rm max}=5~{\rm kpc}$. The column density is then computed by assigning to each voxel in the grid the density of the nearest simulation cell and numerically integrating radially, assuming piecewise-constant values along each ray \citep[as in][]{bennett+_2024}. For the second method, we sample $20000$ viewing angles isotropically and explicitly trace through the \arepo{} mesh with \vl{} to compute the column density (see Section~\ref{sec:vorolite}) from the centre until $d_{\rm max}=5~{\rm kpc}$. We have explicitly checked that extending $d_{\rm max}$ up to 50~kpc does not significantly alter our conclusions for either method, as most of the dense gas is confined to the centre. Note that, for the \fable{}(-sE) runs, the centre is defined as the position of the central SMBH, while for the NoAGN run, we use the halo centre determined from the \textsc{Subfind} halo finder.

Fig.~\ref{fig:hydrogen_column_density} shows the time evolution of the hydrogen column density within $d_{\rm max}=5~{\rm kpc}$. In the NoAGN run, the column density gradually increases with time as large quantities of gas are funnelled from the cosmic web and accumulate in the centre, which cannot be pushed out and regulated by supernova feedback alone. By $z\lesssim7$, most lines of sight appear Compton-thick, highlighting an extreme level of obscuration. This is to be contrasted with simulations with black hole physics, which, on average, have lower values of $N_{\rm H}$ (up to $\gtrsim2~{\rm dex}$ at $z=5.75$) while exhibiting significant time variation due to AGN feedback, which redistributes and ejects gas from the central galaxy. Nonetheless, the simulations predict that the black hole undergoes obscured growth for much of its history, which we discuss in more detail below.

In the lower panel of Fig.~\ref{fig:hydrogen_column_density}, we compare the approximation used in \citet[][]{bennett+_2024} against tracing through the mesh with \vl{}. While their original conclusions are largely unchanged, we find that accurately ray-tracing through the \arepo{} mesh generally yields \textit{higher} column densities, with differences of the order of $5-10$ per cent, and up to $\sim 50$ per cent during certain evolutionary phases. 

In Fig.~\ref{fig:obscuration_detail}, we present a detailed view of various obscuration stages for the central SMBH in the \sE{} simulation. The upper panel reproduces the curve shown in Fig.~\ref{fig:hydrogen_column_density}, including, for alternate snapshots, the full range (min-max) of column densities as probed by 20000 isotropically sampled sightlines. In the lower panels, we show density-weighted radial velocity (upper row) and surface density (lower row) maps of the central $20~{\rm kpc}~\times~20~{\rm kpc}$ at three epochs with distinctive obscuration: A ($z\approx7.59$, obscured), B ($z\approx6.66$, in transition) and C ($z\approx5.75$, unobscured). We also provide a zoomed-in view of the surface density of the inner $2~{\rm kpc}~\times~2~{\rm kpc}$, with the position of the central black hole shown by a black dot (for visual purposes only). In every inset, the dashed circle represents the physical extent within which the rays are computed (5~kpc). 

When obscured (phase A), we see large quantities of inflowing gas surrounding a very dense nucleus, which completely obscures the central engine. At later times (phase B), the gas is distributed irregularly, showing cavities and various gas streams extending outward. In this phase, AGN feedback drives cycles of small blow-out events in the centre, which clear out some gas while leaving some of it behind. This results in a highly anisotropic distribution of gas, wherein some lines of sight are cleared while others remain obscured. In fact, the width of the distribution of column density along the various sightlines remains large, with column densities that cover $\lesssim3-4$~dex from one sightline to another. In particular, there are always some lines of sight that are Compton thick (cf. the grey dashed vertical line), which would signal heavily obscured accretion down to $z\lesssim6$. By $z=5.75$, the immediate surroundings of the black hole are dominated by strong outflows, which have removed gas and significantly lowered the gas density, leaving the engine entirely unobscured (phase C).

The results presented here are consistent with observations and simulations from the literature, which predict that heavily obscured AGN are prevalent at high redshift \citep[e.g.,][]{Vito+_2018, satyavolu+_2023a_obs, Bulichi+_2026}. Specifically, we find that most of the obscuration arises from the dense ISM of the galaxy, with large variations in column density depending on the viewing angle \citep[][]{Circosta+_2019, ni+_2020, gilli+_2022}. Simulations also predict that the high levels of obscuration significantly affect the visibility of SMBHs in the UV, essentially rendering them `invisible' to most wide-field surveys typically used to detect them \citep[e.g.,][]{Trebitsch+_2019, ni+_2020, Vito+_2022, satyavolu+_2023a_obs}. For instance, \citet[][]{bennett+_2024} find that the SMBH in \sE{} would most likely not be detected by present surveys until $z\lesssim7$, after which feedback from the quasar clears out some line of sights that could be detected, with the estimated probability of detection peaking at $\approx50~{\rm per~cent}$ by $z=5.75$.

\subsection{Hot and cold outflowing gas}

\begin{figure}
    \centering
    \includegraphics[width=\linewidth]{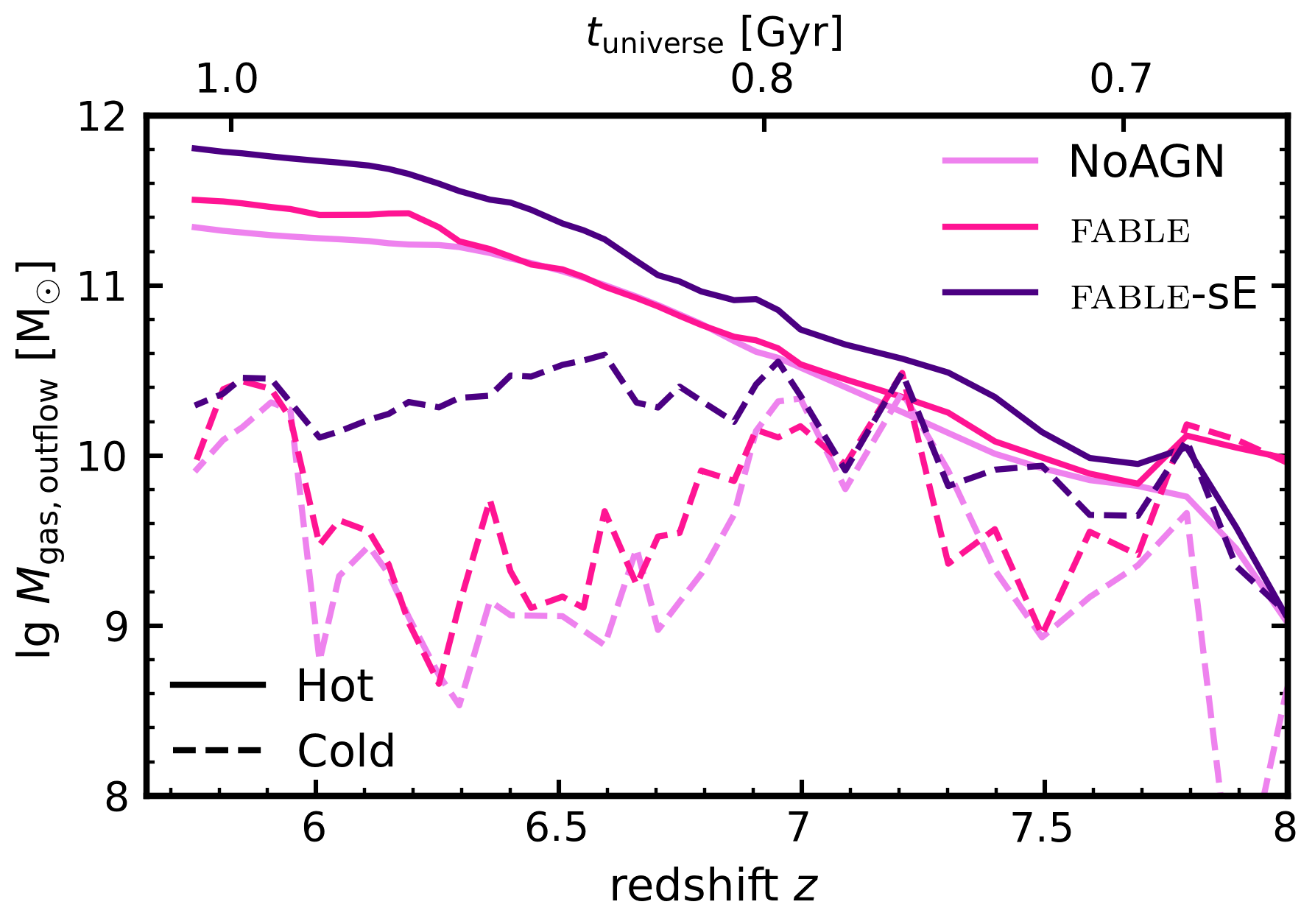}
    \caption{Outflowing gas mass as a function of redshift for each simulation used in this work. All gas within a sphere of radius $3~\rvir$ centred on the SMBH and radial velocity $v_{\mathrm{rad}}\geq300~{\rm km~s^{-1}}$  is considered here. Hot ($T\geq10^6~{\rm K}$) gas is indicated by solid lines, while cold ($T\leq5\times10^4~{\rm K}$ + star-forming) gas is indicated by dashed lines. \emph{In the \sE{} run, there is generally more outflowing gas at all times than in other runs as a result of strong feedback. Cold gas is ubiquitous and sustained in the outflows for the \sE{}, whereas the NoAGN and \fable{} simulations see large variations in outflowing cold gas mass. At $z\lesssim7$, we find over one dex difference in outflowing cold gas mass as a result of a strong blow-out event lasting $\approx200~{\rm Myr}$ in the \sE{} run.}}
    \label{fig:outflows_mass}
\end{figure}

\begin{figure*}
    \centering
    \includegraphics[width=\linewidth]{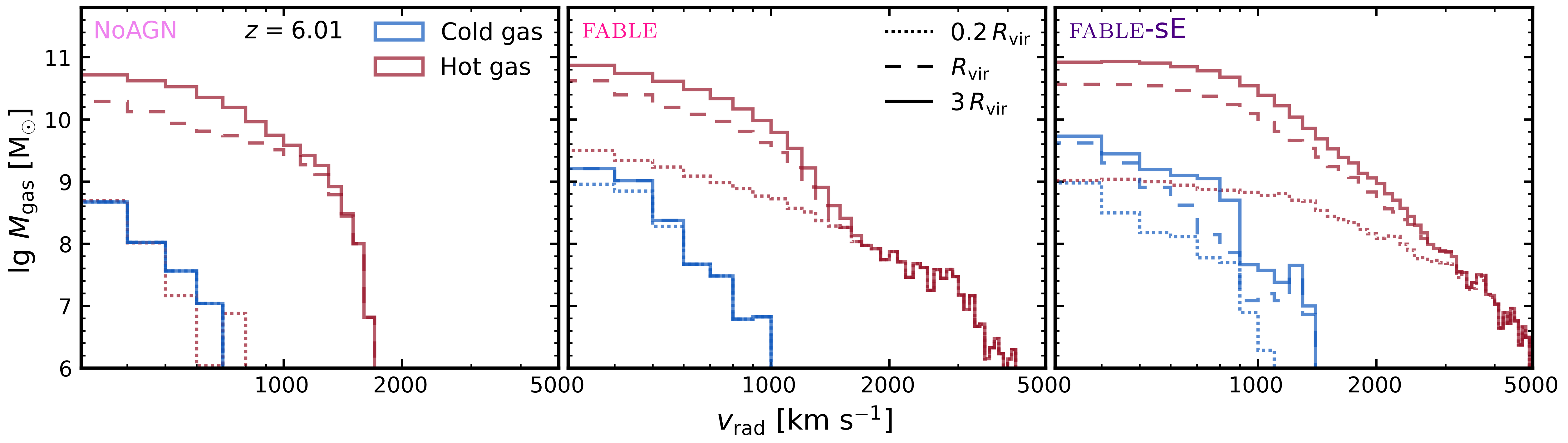}
    \caption{Outflowing gas mass per radial velocity bin within a sphere of varying radius $R_{\mathrm{max}}$ centred on the SMBH at redshift $z=6.01$. Each panel shows results, from left to right, for the NoAGN, \fable{} and \sE{} simulations. Hot (cold) gas is shown by the reddish (bluish) line histograms; the dotted, dashed and continuous lines indicate $R_{\mathrm{max}} \, / \, \rvir=0.2, 1, 3$, respectively. \emph{Gas in the \sE{} run is consistently found at higher radial velocities. In addition, the more abundant cold, outflowing gas reaches velocities of up to $v_{\mathrm{rad}}\gtrsim1300 \, \mathrm{km \ s^{-1}}$, and is present up to and beyond $\gtrsim\rvir$ in the CGM and the IGM of the halo.}}
    \label{fig:outflows_radius}
\end{figure*}

We now focus on the outflow properties in the simulations. We adopt a similar convention as introduced in \citet[][]{costa_sijacki_haehnelt_2015}, and distinguish two phases: `hot gas' with $T\geq10^6~{\rm K}$ and `cold gas' with $T\leq5\times10^4~{\rm K}$, further including star-forming gas. Additionally, we select outflowing gas using a velocity threshold of $v_{\rm rad}\geq300~{\rm km \ s^{-1}}$, to search for AGN-driven outflows. The broad velocity distribution of gas accelerated by AGN makes it challenging to isolate outflows due to AGN feedback from motions caused by stellar feedback or gravity \citep[][]{costa_sijacki_haehnelt_2015, ward+_2024}. We verified that using a more conservative velocity cut of $v_{\mathrm{rad}}\geq500 \, \mathrm{km \ s^{-1}}$ deepens the differences between the \sE{} and the other simulations, strengthening our conclusions.

In Fig.~\ref{fig:outflows_mass}, we show the total outflowing mass within $3~\rvir$ for each simulation, distinguishing hot (solid curves) and cold gas (dashed curves), as a function of redshift. Different colours represent the different simulations.
Outflows have substantial cold and hot components at all times, though hot outflows always dominate the mass budget. The outflowing masses are initially comparable, but significantly diverge at $z<7$, after which the hot component dominates by over one order of magnitude and the cold gas is much less abundant. We find relatively small differences in the outflow masses between the NoAGN and \fable{} simulations, with a slight increase in the hot outflowing gas for $z\leq6.3$ in \fable{}, suggesting that feedback from the massive black hole in \fable{} can affect the CGM, though it is delayed and less powerful. This is to be contrasted with the \sE{} simulation, wherein there is more hot \textit{and} cold outflowing gas at all times, indicating that this simulation produces systematically stronger feedback. We also note that the presence of cold gas in outflows is only ubiquitous in the \sE{} simulation. At $z\lesssim7$ in the \sE{} run, the repeated effect of small-scale feedback injection by the AGN, as discussed in Section~\ref{sec:obscuration}, culminates in a strong blow-out event. This results in a massive ($\sim10^{10}~\msun$) and sustained ($\approx200~{\rm Myr}$) cold gas outflow that sweeps through the halo, approximately one dex more massive than in the other simulations. 

We investigate the radial extent and kinematic properties of the outflowing gas in Fig.~\ref{fig:outflows_radius}. The panels show histograms of the mass of outflowing gas binned by radial velocity. From left to right, we show results for NoAGN, \fable{}, and \sE{}. As in Fig.~\ref{fig:outflows_mass}, we separate hot and cold phases, shown in red and blue, respectively. Different line styles show the distributions obtained within spheres of increasing radius $R_{\mathrm{max}} \, / \, \rvir=0.2, 1, 3$. Signatures of AGN feedback are evident in both \fable{} and \sE{}, where hot gas is ejected to exceptionally high velocities ($v_{\mathrm{rad}}\gtrsim3000-5000~{\rm km~s^{-1}}$). We also see striking differences in the properties of the cold outflowing gas in the \sE{} run compared to the other two. There are large quantities ($\gtrsim5\times10^9~\msun$) of cold gas reaching far into the CGM ($\lesssim\rvir$) \textit{and} IGM (up to $\approx 3~\rvir$), whereas the majority of cold gas in the NoAGN and \fable{} simulations remains in the central galaxy and is confined to the inner halo ($\leq0.2~\rvir$). Additionally, cold gas in \sE{} is very fast, with a significant amount ($\gtrsim 10^8~\msun$) with $v_{\mathrm{rad}}>1000~{\rm km~s^{-1}}$, reaching up to $v_{\mathrm{rad}}\approx 1300~{\rm km~s^{-1}}$.

\begin{figure*}
	\includegraphics[width=\linewidth]{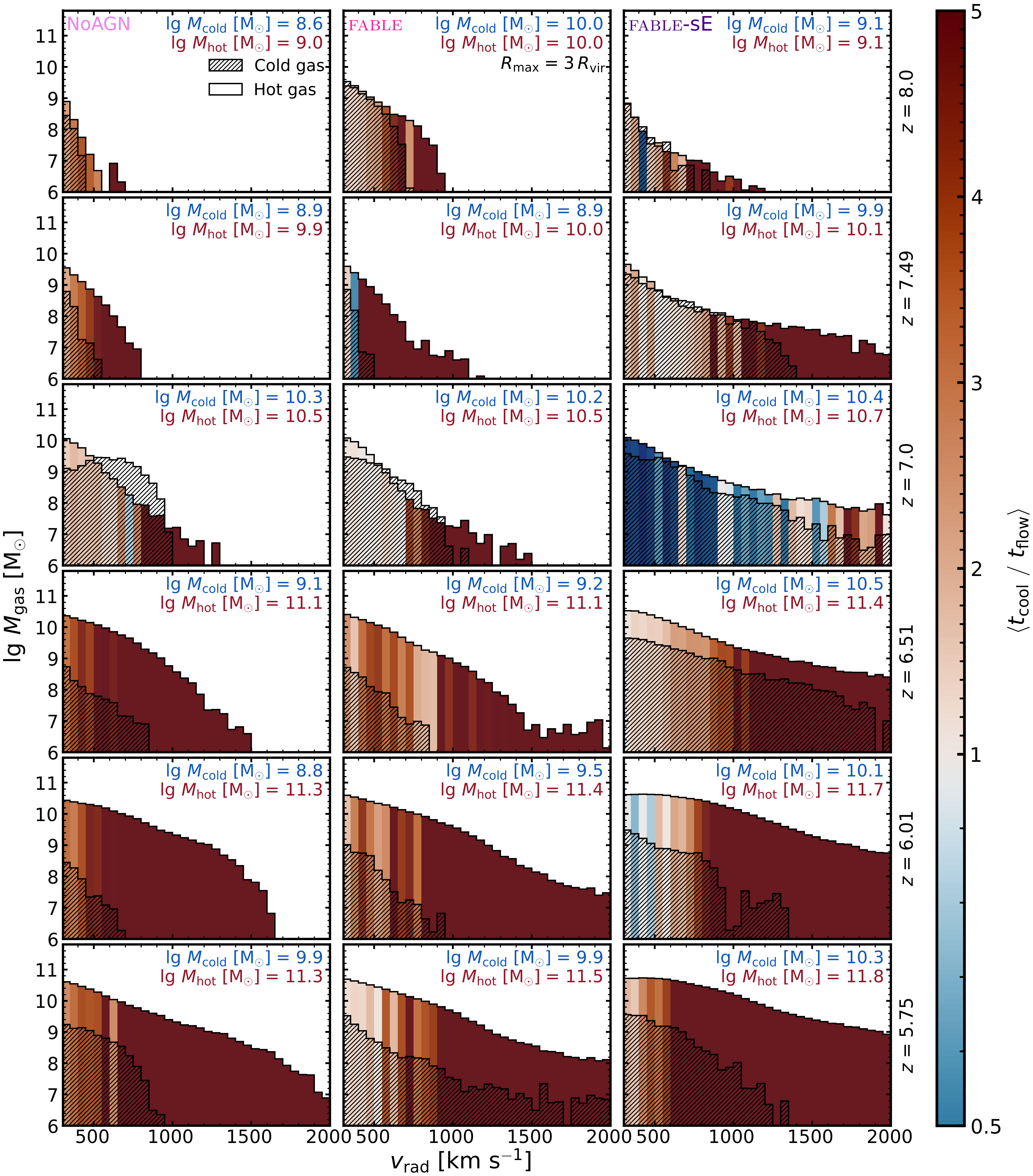}
    \caption{Outflowing gas mass per radial velocity bin within a sphere of radius $3\,\rvir$ centred on the black hole. The left, middle and right columns correspond to results for the NoAGN, \fable{} and \sE{} run, respectively, while the rows show histograms for redshifts $z=8, 7.5, 7, 6.5, 6$ and $5.75$, from top to bottom (as also indicated on the right-hand side of every row). Hot (cold) gas is displayed with empty (filled) histograms. The total mass of the hot and cold outflowing gas is shown in the upper-right corner of every panel. 
    Radial velocity bins for the hot gas are colour-coded by the density-weighted ratio of cooling to flow time $\langle t_{\rm cool} \, / \, t_{\rm flow}\rangle$ according to the colourbar on the right, such that $t_{\mathrm{cool}} > t_{\mathrm{flow}}$ is in red and $t_{\mathrm{cool}} < t_{\mathrm{flow}}$ is in blue. \emph{Larger amounts of fast, cold outflowing gas are found in the \sE{} run, traced in part to the efficient cooling of hot gas in the outflows.}}
    \label{fig:outflows_redshift}
\end{figure*}

We now analyse the properties of the hot phase to understand whether it can efficiently cool to give rise to the significant cold component we find in the outflows \citep[][]{costa_sijacki_haehnelt_2015}. Outflow cooling is expected in regions where the cooling time is shorter than the flow time. We define the cooling time $t_{\mathrm{cool}}=u_{\rm int}\,/\,(n_{\rm H}^2\,\Lambda_{\rm cool})$, where $u_{\rm int}$ is the internal energy, $n_{\rm H}$ is the gas density and $\Lambda_{\rm cool}$ is the net cooling rate; and the flow time is $t_{\rm flow}=r\,/\,v_{\rm rad}$, where $r$ is the radial position and $v_{\rm rad}$ is the radial velocity. In Fig.~\ref{fig:outflows_redshift}, we show histograms of the outflowing gas mass binned by radial velocity for the different simulations (columns) and as a function of redshift (decreasing, from top to bottom). The cold gas is indicated by line-filled histograms, whereas the distribution of hot gas is colour-coded by the density-weighted ratio of cooling to flow time $\langle t_{\mathrm{cool}}\,/\,t_{\mathrm{flow}}\rangle$. Cooling of the hot outflowing component occurs when this ratio is $\leq1$, shown in bluish hues.

The kinematics of both hot and cold gas for the \fable{} and \sE{} runs differ from the NoAGN simulation at all redshifts. At higher redshifts ($z\sim 8$), a large fraction of the hot outflow is in the rapid cooling regime (bluish hues), which explains why the masses of the two components are initially similar. However, the hot gas cannot cool as efficiently at later times, thus explaining why there is more hot than cold gas at lower redshifts. At $z\lesssim7$, we see that there are large quantities of hot gas with short cooling times only in the \sE{} run. We identify this as the primary mechanism for the formation and sustenance of fast, cold gas in the outflows in \sE{}, which is only ubiquitous in that simulation and can at times reach extreme velocities ($v_{\rm rad} \gtrsim 2300~{\rm km~s^{-1}}$). In particular, we note such rapidly-cooling hot gas is not found in the NoAGN or the \fable{} simulations, thus explaining how the $\sim1$~dex difference in outflowing cold mass arises (cf. Fig.~\ref{fig:outflows_mass}). By $z<6$, there is very little gas that can efficiently cool in the hot outflows of the \sE{} simulation, and we see that the cold gas masses are again similar across the suite.

\begin{figure*}
	\includegraphics[width=\linewidth]{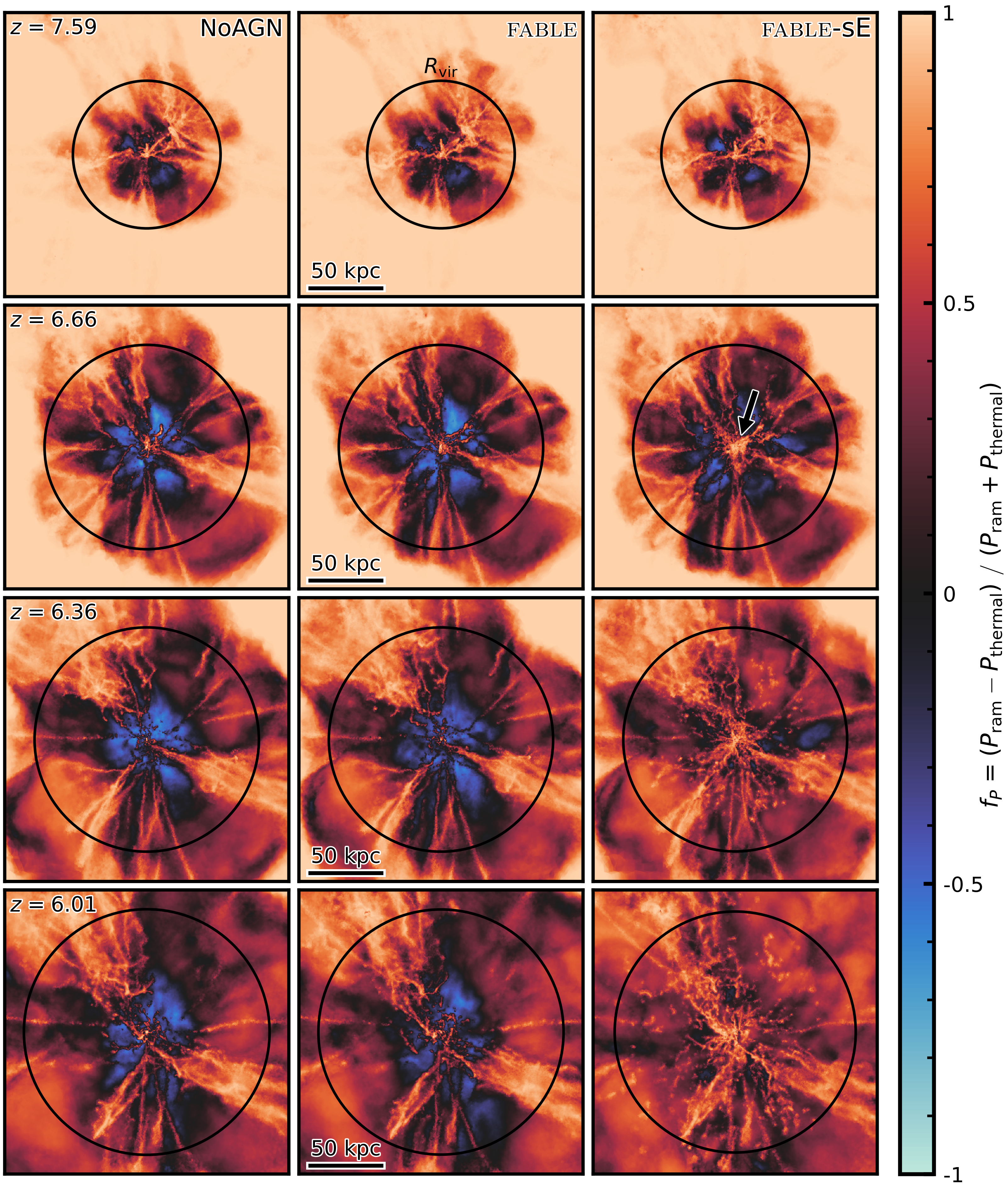}
    \caption{Density-weighted projected pressure ratio $f_P = (P_{\mathrm{ram}} - P_{\mathrm{thermal}}) \, / \, (P_{\mathrm{ram}} + P_{\mathrm{thermal}})$ for varying redshift (rows) for the NoAGN (left column), \fable{} (middle column) and \sE{} (right column) simulation. Red-orange (blue-cyan) hues indicate regions where ram (thermal) pressure dominates. \emph{Prior to the blow-out phase in the \sE{} run ($z\gtrsim7$), the CGM in all three simulations is similar, with gas being in large proportion ram-pressure dominated as it falls into the central halo. In later stages, the effect of the strong feedback in \sE{} drives powerful outflows which compress gas in the diffuse halo and shatter cosmic filaments. The NoAGN and \fable{} simulations, on the other hand, have gas in the diffuse halo predominantly supported by thermal pressure as well as well-developed ram pressure-supported filaments.}}
    \label{fig:pressure_ratio}
\end{figure*}

\begin{figure*}
    \centering
    \includegraphics[width=\linewidth]{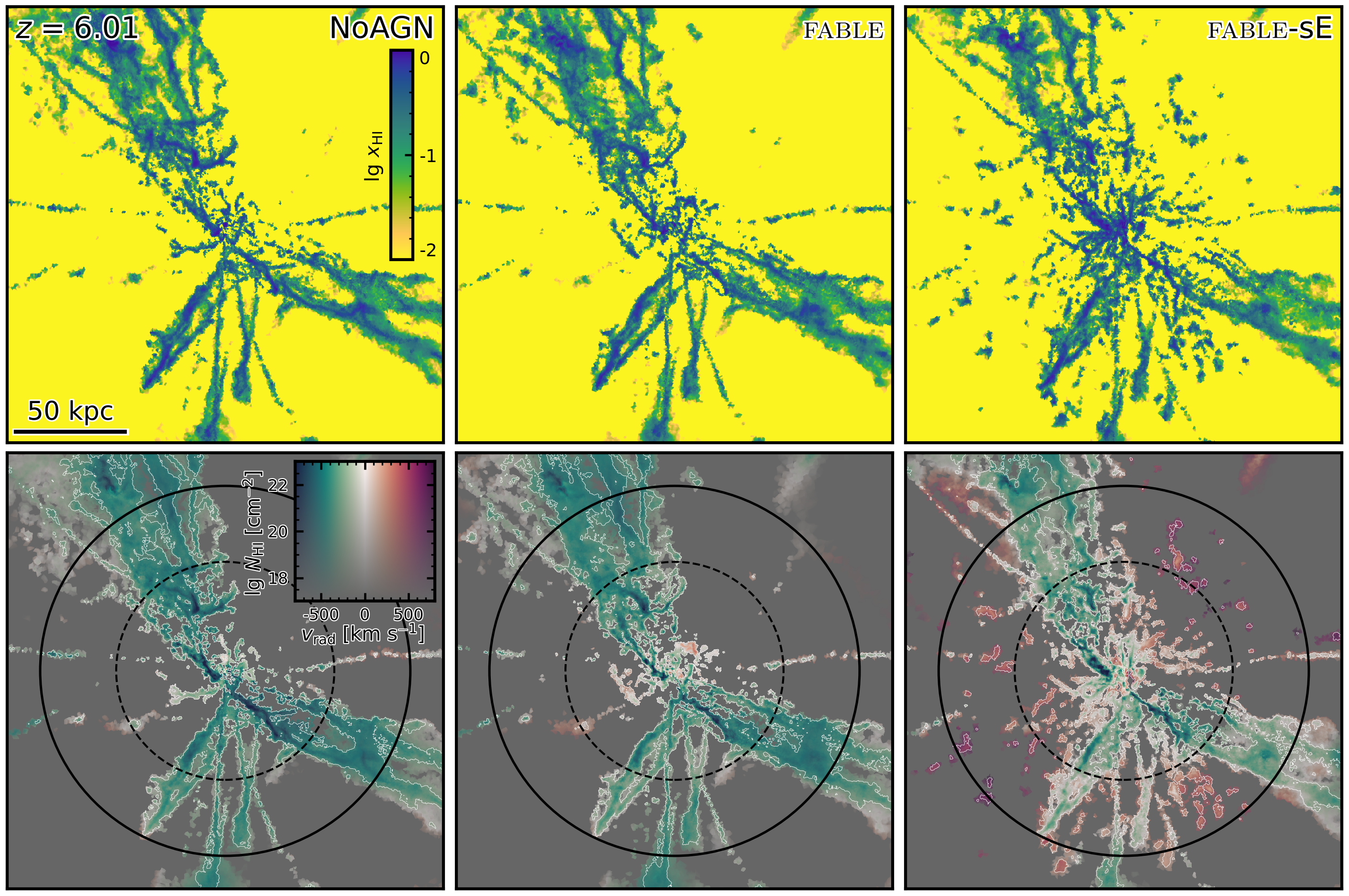}
    \caption{Visual overview of the ionization state of the CGM in the suite of simulations at $z=6.01$. Here, gas is assumed to be in ionization equilibrium with a meta-galactic UVB. From left to right, the different columns show results for the NoAGN, \fable{} and \sE{} simulations, respectively. The upper row displays the density-weighted neutral hydrogen fraction ($x_{\hi{}}$), while the lower row shows 2D bivariate projections of the neutral hydrogen column density ($N_{\hi{}}$) and density-weighted radial velocity ($v_{\rm rad}$). This visualisation enables the simultaneous tracing of outflows and dense, neutral gas. The bivariate colourmap is shown as an inset in the bottom left panel, with the brightness/saturation reflecting the column density and the hue displaying the radial velocity. White contours delineate DLA sightlines ($N_{\hi{}}>10^{20.3}~{\rm cm^{-2}}$). In the lower panels, the full and dashed circles display the virial radius and $R=50$~kpc, respectively. \emph{A large fraction of the CGM in the hot, diffuse phase is ionized by the UVB. Along the cosmic filaments and in the central galaxy, we have $x_\hi{}>0.1$ such that the column density is very high. We note that many more lines of sight are covered by dense and neutral gas in the \sE{} simulation, up to and beyond the virial radius. This excess of gas compared to the two other runs is almost entirely made up of outflowing gas clumps that arise from strong AGN feedback in \sE{}.}}
    \label{fig:f_hi_visual}
\end{figure*}

We continue the investigation into the properties of the outflows in the simulations by describing their interaction with the CGM. In Fig.~\ref{fig:pressure_ratio}, we show maps of the projected density-weighted normalised pressure ratio of gas, $f_P = (P_{\mathrm{ram}} - P_{\mathrm{thermal}}) \, / \, (P_{\mathrm{ram}} + P_{\mathrm{thermal}})$, for the three simulations (columns) and at various redshifts (decreasing, from top to bottom row). This ratio quantifies the dominant source of pressure, differentiating between ram pressure ($P_{\mathrm{ram}}=\rho\,v^2$) and thermal pressure ($P_{\mathrm{thermal}}=n\,k_{\mathrm{B}}\,T$). It can be used to identify regions where gas is dynamically compressed (e.g., by outflows driven by AGN feedback) and to inform on the dominant form of energy (kinetic versus thermal) and Mach number of the gas \citep[e.g.,][]{bourne_sijacki_2017}.

At high redshift ($z>7.5$, upper row), all simulations look qualitatively very similar, with gas on the largest scales being ram pressure-dominated in the IGM and across the filaments which feed gas into the central halo.  Along the filaments, the rapidly accreting cold gas is likely supersonic and ram-pressure dominated as it penetrates through the halo down to the central galaxy, while in the halo we see a mixture of diffuse hot halo gas thermalised approximately at the virial temperature \citep[e.g.,][]{birnboim_dekel_2003} and outflowing, pressure-dominated gas. In the second row for the \sE{} run, the maps show a clear ram pressure-dominated gas region, highlighted with an arrow, growing from the centre and extending outwards. This coincides exactly with a strong blowout event, which happens around $z\lesssim7$. By $z\sim6$, there is a remarkable difference between the \sE{} and other simulations, whereby the gas is ram pressure-dominated in the entire halo. We directly attribute this to the outflows launched from the centre of the halo by the more massive black hole in this simulation. Gas along the filaments is disrupted as they are being shattered, and the gas in the diffuse halo is shocked by the material launched by the AGN. 

Specifically, the highlighted gas clumps in Fig.~\ref{fig:visual_overview} spatially correlate with regions in the diffuse halo that are strongly ram pressure-dominated. This indicates that the formation of cool gas follows from interactions between the AGN-driven outflows and gas in the CGM. This is consistent with detailed simulations of the interaction between AGN winds and a clumpy ISM, which found that the existence and survival of cold gas is linked to the formation of cold clouds by radiative cooling entrained within a faster, hot phase \citep[see, e.g.,][]{Gronke_Oh_2018, Fielding_Bryan_2022, Jennings+_2023, ward+_2024, ward+_2025, Gronke_Schneider_2026}. Our results are also consistent with previous work studying outflows in a cosmological setting at high redshift \citep[][]{costa_sijacki_haehnelt_2015, Biernacki_Teyssier_Bleuler_2017, Biernacki_Teyssier_2018}, in which the presence of cold and dense gas in the outflows is attributed to the combined effect of supernovae and AGN feedback. Specifically, supernovae pre-enrich the CGM with dense and metal-enriched clumps that are swept by the shocked hot gas launched by the AGN. This interaction leads to efficient cooling of the gas, and results in large masses of fast, cold gas in outflows that reach beyond the galaxy and into the CGM and IGM of the host halo.

\subsection{The ionization state of the CGM}\label{sec:ioni-state}

In this section, we inspect the response of the CGM to the radiative activity of the central SMBH in the simulations by considering the ionization state of neutral hydrogen \hi{}.

\subsubsection{Neutral hydrogen in ionization equilibrium with the UVB}\label{sec:UVB-results}

To begin with, we consider the situation in which the gas is in ionization equilibrium with background sources, modelled as a homogeneous, redshift-dependent ionizing UV background. For this investigation, we compute the predicted ionization state of hydrogen using the temperature and density of every gas particle from the hydrodynamical simulations and assume that gas is in equilibrium with the UVB (see details in Section~\ref{sec:methods} and Appendix~\ref{app:neutral-hydrogen}). No additional radiative transfer simulations are performed here. We investigate the effects of quasar radiation in the following Section~\ref{sec:quasar-rad}.

We first provide a visual inspection of simulations at $z=6$ in Fig.~\ref{fig:f_hi_visual}. The upper row shows the ionization state of hydrogen for each simulation in the suite. The majority of the hot, diffuse gas in the halo is fully ionized, while cold, dense gas found in cosmic filaments and in gas clumps is neutral. In the \sE{} run, these clumps are more abundant and cover more lines of sight. In the lower row, we show a bivariate projection to simultaneously trace the velocity (green hues for inflowing and red hues for outflowing gas) and the column density of neutral hydrogen (higher brightness translates to higher column). Along lines of sight with $x_{\hi{}}>0.1$, the column density is very high ($N_{\hi}>10^{17}~{\rm cm^{-2}}$), with a large fraction of sightlines in DLA systems with $N_\hi{}>10^{20.3}$\,cm$^{-2}$ (highlighted as white contours). The neutral gas clumps populate the CGM of the halo in \sE{}, resulting in multiple lines of sight covered by high column densities, which are almost entirely found in the outflows (orange/red hues). These are not present in either the NoAGN or \fable{} runs, highlighting that it is a consequence of the strong AGN feedback in the \sE{} simulation driven by the early assembly of an extremely massive black hole. 

\begin{figure}
    \centering
    \includegraphics[width=\linewidth]{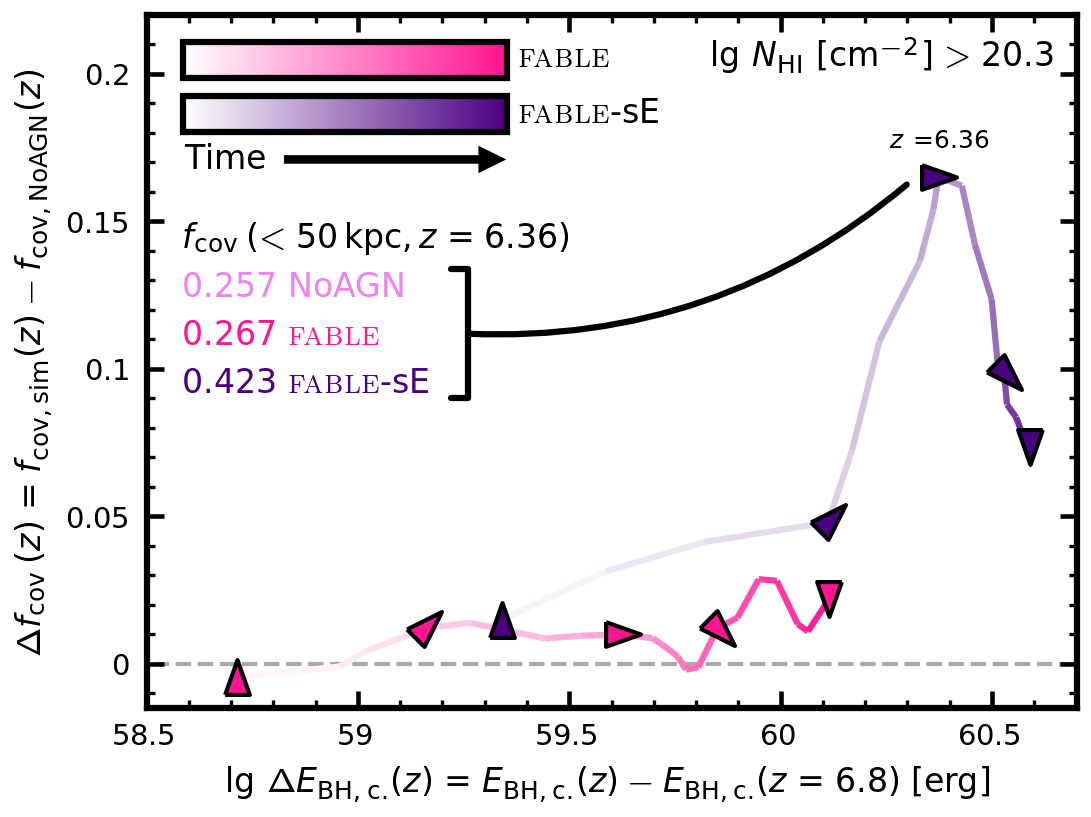}
    \caption{Difference in the \hi{} covering fraction within 50~kpc between the NoAGN run and the \fable{} and \sE{} runs versus cumulative feedback energy injected by the black hole in quasar mode. We adopt the threshold for DLA systems ($N_{\hi{}} > 10^{20.3}$\,cm$^{-2}$) for the covering fraction. For the abscissa, we compute the difference in cumulative energy injected by the accreting black hole into its surroundings, using the value at $z=6.8$ as the zero-point for each simulation. The coloured lines show tracks through time as encoded in the legend (lighter colours indicate earlier times/higher redshift), and we use markers to match specific snapshots across the simulations. We highlight $z=6.36$ when the difference in covering fraction is highest. \emph{At redshift $z=6.6$, there is already one dex more energy injected by the black hole in \sE{} than in the \fable{} run. With the blow-out developing, $\Delta f_{\rm cov}$ increases as strong outflows push dense, neutral clumps to cover a larger area. The covering fraction increases by up to 64 per cent, before decreasing once the gas goes beyond 50~kpc and up to $\gtrsim \rvir$.}}
    \label{fig:delta_fcov_delta_Ebh}
\end{figure}

We thus see a connection between the strength of outflows originating from AGN feedback and the \hi{} covering fraction. This is related to the blow-out in \sE{}, during which large quantities of dense, cold and neutral gas (clumps) are expelled, intercepting more lines of sight out to $\sim\rvir$. For any given radius $R$, the covering fraction of neutral hydrogen $f_{\rm cov}\, (<R)$ is expected to increase as the clumps fill the area, then reach a peak and finally decrease once the clumps have left the region. Naturally, the timing and exact value for these changes will depend on the chosen impact parameter within which the covering fraction is evaluated.

We investigate the variation of the DLA covering fraction in Fig.~\ref{fig:delta_fcov_delta_Ebh}. We compare $f_{\rm cov}(<R=50\,{\rm kpc})$\footnote{We have checked that these conclusions are not affected by choosing $R\gtrsim0.2\rvir\approx 17$~kpc at $z=6$, where most of the difference is found, see Fig.~\ref{fig:outflows_radius}.} for DLAs versus the cumulative energy injected by the accreting black hole. In the ordinate, we show $\Delta f_{\rm cov}(z) = f_{\rm cov, \, sim}(z) - f_{\rm cov, \, NoAGN}(z)$, the difference in the covering fraction within 50~kpc between the NoAGN and the \fable{} and \sE{} runs, respectively. In the abscissa, we show $\Delta E_{\rm BH, \, cumulative} (z) = E_{\rm BH, \, cumulative} (z) - E_{\rm BH, \, cumulative}(z= 6.8)$, which corresponds to the difference in injected energy by the black hole from the beginning of the blow-out phase in \sE{} $(z=6.8)$. We show the evolution of both quantities for the \fable{} and \sE{} simulations with coloured tracks. The time evolution is indicated by the transparency of the lines, starting from $z=6.8$ (lighter colours) down to $z=5.75$ (darker colours). We also use markers to help match the same snapshots for the simulations.

Focusing first on \fable{}, we note very little difference in the DLA covering fraction from the NoAGN run at all times, peaking at around $\Delta f_{\rm cov}\approx0.03$. The large quantities of energy injected by the accreting black hole do not result in outflows which quantitatively distinguish the \hi{} content in its CGM from that of a halo without any black hole. For the \sE{} run, we note that at $z=6.6$ (marker pointing up and to the right) there is already around one dex more injected energy by the central SMBH than in \fable{}, and a steady rise in $\Delta f_{\rm cov}$. The disruption caused by the SMBH is then clearly visible as the enhancement of the covering fraction within 50~kpc reaches a maximum of $\Delta f_{\rm cov}\approx0.17$ at $z=6.36$, during the blow-out phase. We also show the DLA covering fraction value within 50~kpc at that time in the figure. The enhancement corresponds to a $\approx 64$ per cent increase with respect to the NoAGN covering fraction of DLAs, which can be attributed to the presence of cold, dense clumps of neutral gas which cover more sightlines. As noted previously, this is a transient phenomenon, as the blow-out phase ejects gas for $\approx200$~Myr; by the end of the simulation ($z=5.75$), there is a noticeable but more modest enhancement ($\Delta f_{\rm cov}\approx0.07$) as most clumps reach $R>50$~kpc
\footnote{We verified that the conclusions are not dependent on the chosen line of sight \citep[][]{tortora+_2024, gelli+_2025}, as we find enhancements of $\Delta f_{\rm cov}\approx79$ per cent and $\Delta f_{\rm cov}\approx50$ per cent along two further orthogonal projection axes for the \sE{} run at $z=6.36$.}. 

\begin{figure}
    \centering
    \includegraphics[width=\linewidth]{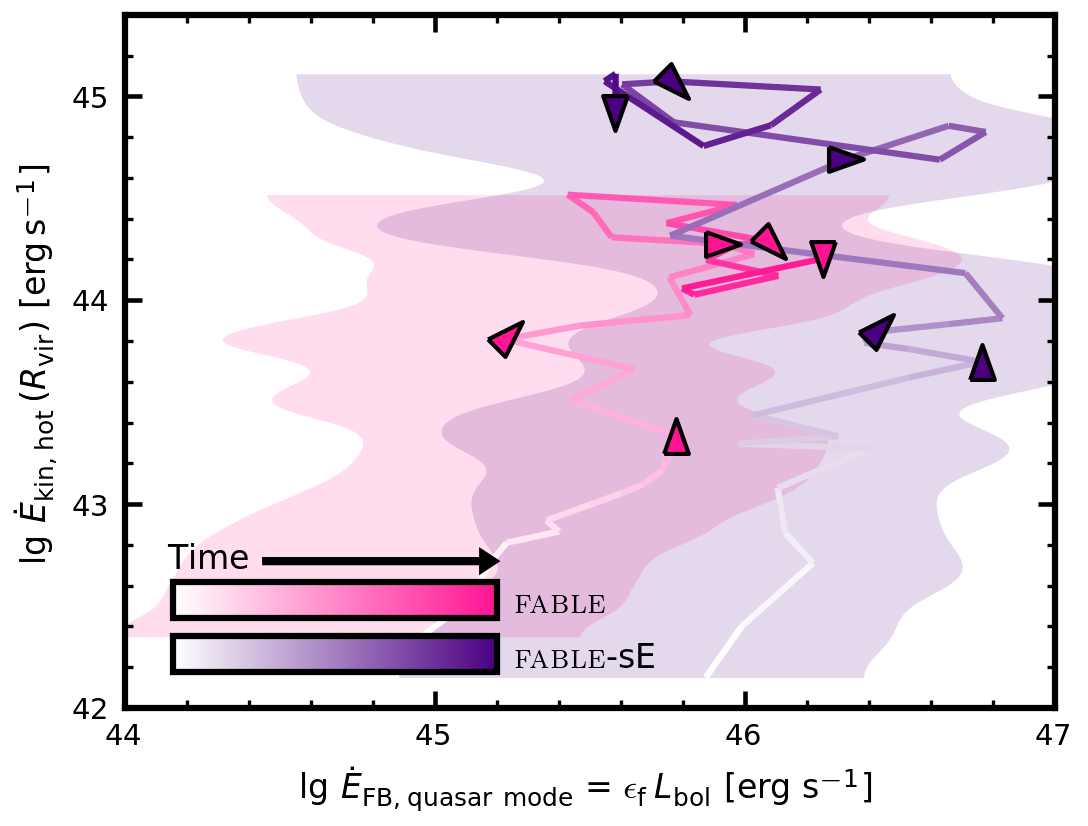}
    \caption{Kinetic energy outflow rate of hot, outflowing gas $\dot{E}_{\mathrm{kin, \, hot}}$ versus kinetic feedback energy from the AGN in quasar mode feedback $\dot{E}_{\mathrm{FB, \, quasar \, mode}}$. The coloured lines show tracks through time as encoded in the legend (lighter colours indicate earlier times/higher redshift, starting at $z=7.5$), and we use markers to match specific snapshots across the simulations (as in Fig.~\ref{fig:delta_fcov_delta_Ebh}). There is a large scatter in the estimate of the bolometric luminosity, which we display with hulls indicating a +0.5/-1 dex variation in $L_{\rm bol}$. \emph{The instantaneous activity of the central SMBH is not directly related to the strength of outflows. At early times, $\dot{E}_{\rm FB}$ is around one dex higher in \sE{} while $\dot{E}_{\rm kin}$ remains the same. By the end of the simulation ($z=5.75$, downward triangle), $\dot{E}_{\rm FB}$ is five times higher in \fable{} but $\dot{E}_{\rm kin}$ is four times lower than in \sE{}. Strong outflows originate from the cumulative effect of AGN feedback injecting large amounts of energy into its surroundings over extended periods of time.}}
    \label{fig:Ekin_Efb}
\end{figure}

The enhancement in the quantity of neutral gas is a result of strong outflows launched by AGN feedback. These outflows are the result of \textit{cumulative} energy injection by the accreting SMBH at the centre of the halo (see abscissa in Fig.~\ref{fig:delta_fcov_delta_Ebh}), which disturbs and expels gas in its vicinity and creates channels out of which large outflows may escape after some time. In particular, we note that outflows may not be connected to the \textit{instantaneous} activity of the central engine, which we show in Fig.~\ref{fig:Ekin_Efb}, and caution against directly correlating the activity of the black hole with the presence of outflows. In Fig.~\ref{fig:Ekin_Efb}, we quantify this claim by comparing a measure of outflow strength with a measure of the black hole's instantaneous activity. For the former we choose the kinetic energy outflow rate of hot gas at the virial radius $\dot{E}_{\mathrm{kin, \, hot}}(\rvir)$ and for the latter we choose the instantaneous thermal feedback energy injected by the SMBH in quasar mode accretion $\dot{E}_{\mathrm{FB, \, quasar \, mode}}$. The kinetic outflow rate is computed by selecting all hot, outflowing gas within a shell at $\rvir$ with thickness $\Delta r=20$~kpc, and summing up the contribution of each cell \citep[as in][]{koudmani+_2021}:
\begin{equation}\label{eq:Ekin_of}
    \dot{E}_{\mathrm{kin}}\,(\rvir)=\frac{1}{2}\sum_{\mathrm{cell} \, \in \, \rvir \pm \frac{\Delta r}{2}} \frac{m_{\mathrm{cell}} \, v_{\mathrm{rad}}}{\Delta r} \, v_{\mathrm{rad}}^2 \, ,
\end{equation}
while the thermal feedback energy is proportional to the bolometric luminosity $L_{\rm bol}$ (or, equivalently, accretion rate $\dot{M}_{\rm BH}$), see equation~\eqref{eq:fb-quasar}.

We show the evolution of both quantities as a function of time in Fig.~\ref{fig:Ekin_Efb}. For the injected feedback energy $\dot{E}_{\mathrm{FB, \, quasar \, mode}}$, the lines are computed from the value of bolometric luminosity computed by taking a rolling mean of width 1000 time-steps around each snapshot. There is also significant scatter, due to large variations in accretion rate from one time-step to another, which we display with hulls indicating a +0.5/-1 dex variation in $L_{\rm bol}$ \citep[][]{bennett+_2024}. The time evolution is indicated by the transparency of the lines, starting at $z=7.6$ down to $z=5.75$, corresponding to the final snapshot of the simulations. We use markers to match snapshots for the simulations, with the same values as in Fig.~\ref{fig:delta_fcov_delta_Ebh}. The two quantities are initially correlated, with increased energy injection from feedback causing stronger outflows, as signalled by the increased kinetic outflow rate measured at the virial radius. This is likely the case because, at higher redshift, the virial radius is smaller, thus causing the outflow crossing time to also be short. We see that the kinetic outflow rate can evolve rapidly or more slowly depending on the cosmic time and simulation, but significantly increases after some time in \sE{}. The kinetic outflow rate and injected feedback energy are also decoupled at later times. This is particularly the case for the \sE{} run, in which we see that the feedback energy varies rapidly by $\sim1$~dex as a result of decreased accretion from self-regulation, while the kinetic outflow rate remains high. For the \fable{} run, both quantities stagnate at later times. 

The feedback energy injection rate thus does not correlate very well with stronger outflows as measured, for instance, with the kinetic energy outflow rate. In particular, at $z<6$, the former is \emph{higher} in the \fable{} run, even though the outflows are stronger in \sE{}. Our simulations clearly show that instantaneous AGN activity cannot be linked to the presence of sustained outflows further out into the halo ($\gtrsim50$~kpc at $z=6$). Instead, outflows are the result of cumulative energy injection by the accreting SMBH at the centre of the halo, which disturbs and expels gas in its vicinity and creates channels out of which large outflows may escape after some time.

Additionally, we caution that there is significant scatter in the abscissa of Fig.~\ref{fig:Ekin_Efb}, because of very large and rapid variations in the instantaneous accretion rate and bolometric luminosity of the black holes in the \fable{} and \sE{} simulations. These rapid variations should be interpreted cautiously, as detailed modelling of gas accretion through a non-steady accretion disc is ultimately needed to robustly quantify and interpret the likely variability. We thus generally caution against directly comparing and correlating the instantaneous activity of the accreting SMBH with the presence of sustained outflows.

\subsubsection{The effects of quasar radiation}\label{sec:quasar-rad}

In this section, we present the results of radiative transfer simulations to assess the effects of quasar radiation in post-processing. The initial state for the radiative transfer simulations corresponds to the predicted ionization equilibrium with the UVB presented in the previous Section~\ref{sec:UVB-results}. We run the simulations for a total time of $t_{\rm{quasar}}\approx1~{\rm Myr}$, which we find to be sufficiently long for the simulations to reach photoionization equilibrium. In this work, the ionizing photon rate is kept constant throughout the radiative transfer, and we leave the study of time-dependent light curves for future work. We focus our attention on two comparisons: across the suite of simulations at $z=6$, and at three key obscuration stages in \sE{}.

\begin{figure*}
    \centering
    \includegraphics[width=\linewidth]{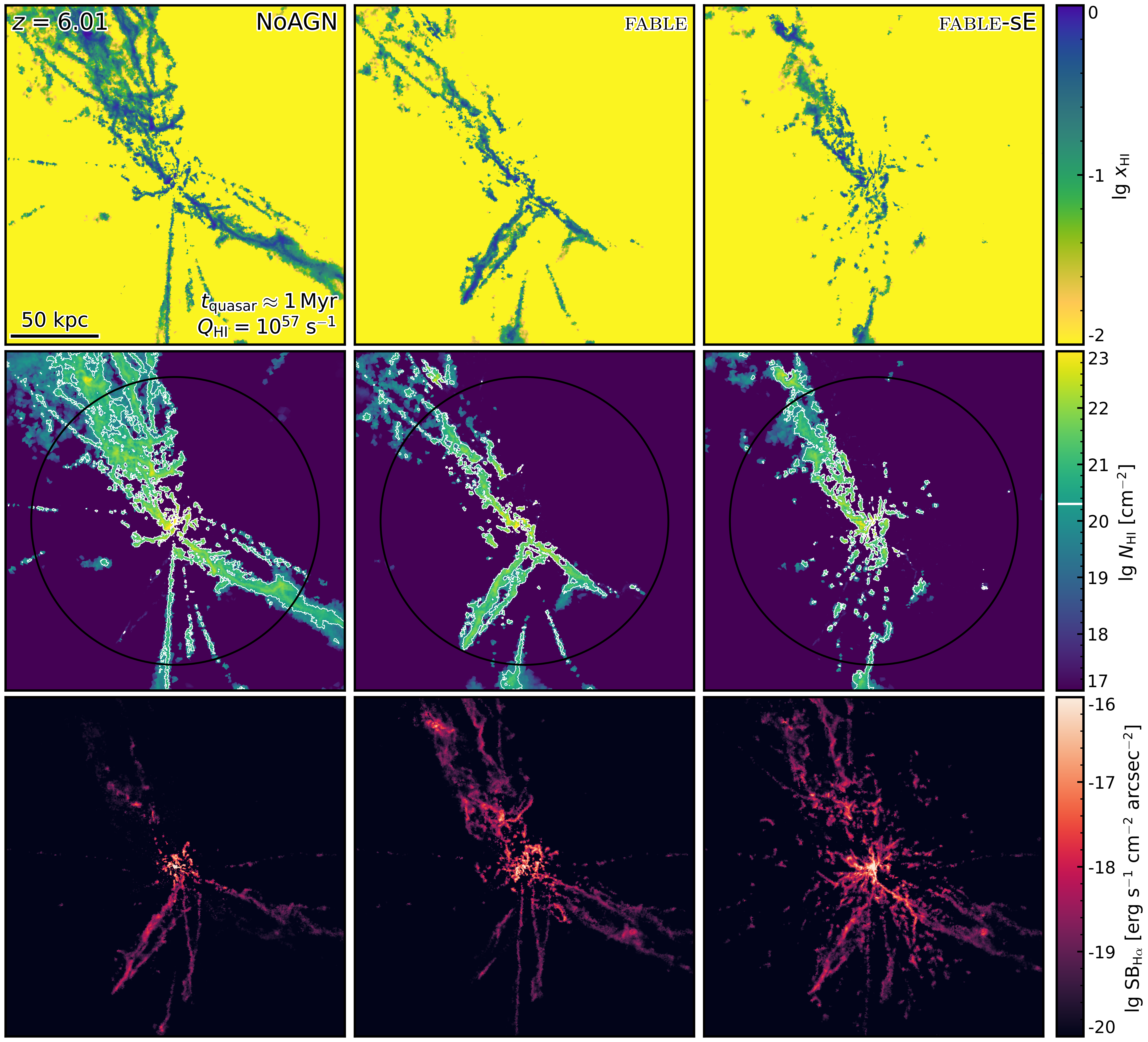}
    \caption{Visual overview of the ionization state of the CGM of the NoAGN (\textit{left column}), \fable{} (\textit{middle column}) and \sE{} (\textit{right column}) runs at $z=6.01$ after a quasar event with a lifetime of around 1 Myr. The constant rate of ionizing photons is chosen to be $Q_{\hi{}}=10^{57}~{\rm s^{-1}}$ for all simulations. We show the density-weighted neutral hydrogen fraction ($x_{\hi{}}$, \textit{upper row}), the neutral hydrogen column density ($N_{\hi{}}$, \textit{middle row}) and $\rm H\alpha$ surface brightness ($\rm SB_{\rm H\alpha}$, \textit{lower row}). White contours delineate DLA sightlines ($N_{\hi{}}>10^{20.3}~{\rm cm^{-2}}$). \emph{The ionization state of the CGM heavily depends on the physical state of the hydrodynamical simulation. In the NoAGN run, the radiation is confined to preferential directions and only reaches certain parts of the CGM, due to high hydrogen column densities in the centre. In \fable{}, the radiation can reach the CGM and IGM of the halo, photoionizing gas along the cosmic filaments. In \sE{}, a much greater portion of the halo is ionized, including the gas clumps and substructures along the shattered cosmic filaments. This results in the formation of more diffuse and extended H$\alpha$ nebulae, which directly probe the effect of strong AGN feedback.}}
    \label{fig:f_hi-sims-snap125}
\end{figure*}

We show the effects of performing the radiative transfer simulations using the same ionizing luminosity in the different simulations of our suite at $z=6.01$ in Fig.~\ref{fig:f_hi-sims-snap125}. Each column shows one of the simulations and the different rows show the neutral hydrogen fraction, neutral hydrogen column density and H$\alpha$ surface brightness. The chosen constant ionizing luminosity is $Q_{\hi{}}=10^{57}$\,s$^{-1}$. This roughly corresponds to the ionizing photon production rate of a typical $L_{\rm bol}\approx10^{47}~{\rm erg~s^{-1}}$ quasar assuming the \textsc{qsosed} family of AGN spectra \citep[][]{kubota_done_2018}. While there is no accreting black hole in the NoAGN run, we also post-process this simulation, assuming the same luminous source is placed at the centre of the halo, to assess the effect of pre-processing by AGN feedback on the escape of radiation. 

\begin{figure*}
    \centering
    \includegraphics[width=\linewidth]{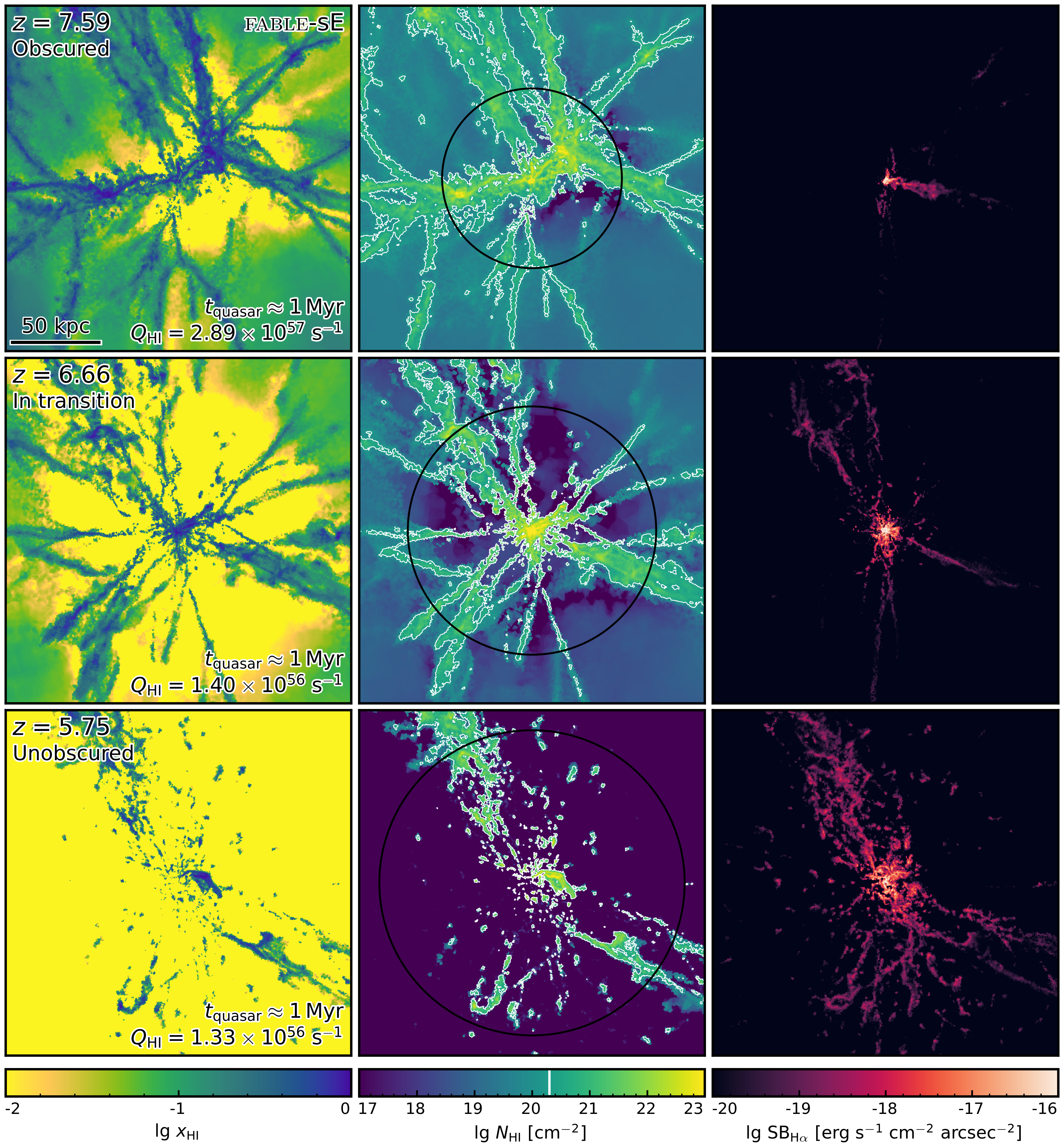      }
    \caption{Visual overview of the ionization state of the CGM of the \sE{} simulation at $z=7.59$ (\textit{top row}), $z=6.66$ (\textit{middle row}) and $z=5.75$ (\textit{lower row}) after a quasar event with a lifetime of around 1 Myr. The redshifts are chosen to correspond to three obscuration stages of the central engine, respectively, `obscured', `in transition' and `unobscured' as highlighted in Fig.~\ref{fig:obscuration_detail}. The rate of ionizing photons is computed from the predicted bolometric luminosity of the quasar at every snapshot with a correction obtained using spectra from the QSOSED model. We show the density-weighted neutral hydrogen fraction ($x_{\hi{}}$, \textit{left column}), the neutral hydrogen column density ($N_{\hi{}}$, \textit{middle column}) and $\rm H\alpha$ surface brightness ($\rm SB_{\rm H\alpha}$, \textit{right column}). White contours delineate DLA sightlines ($N_{\hi{}}>10^{20.3}~{\rm cm^{-2}}$). \emph{When the quasar is obscured, the ionizing radiation cannot escape the central region, and we only see very faint H$\alpha$ emission from one direction, mostly confined to the centre. In the transition stage, some of the radiation can escape along channels that have been cleared out by AGN feedback and reach $R\gtrsim100~{\rm kpc}$, though we note that the $\rm H\alpha$ emission remains centrally concentrated and is anisotropic. As the cumulative effect of AGN feedback removes gas and unobscures the central engine at $z=5.75$, radiation can escape and photoionize gas deep in the CGM and IGM, igniting both cosmic filaments and clumps in H$\alpha$ despite lower ionizing photon rates from the source.}}
    \label{fig:f_hi-sE-obs}
\end{figure*}

\begin{figure*}
    \centering
    \includegraphics[width=\linewidth]{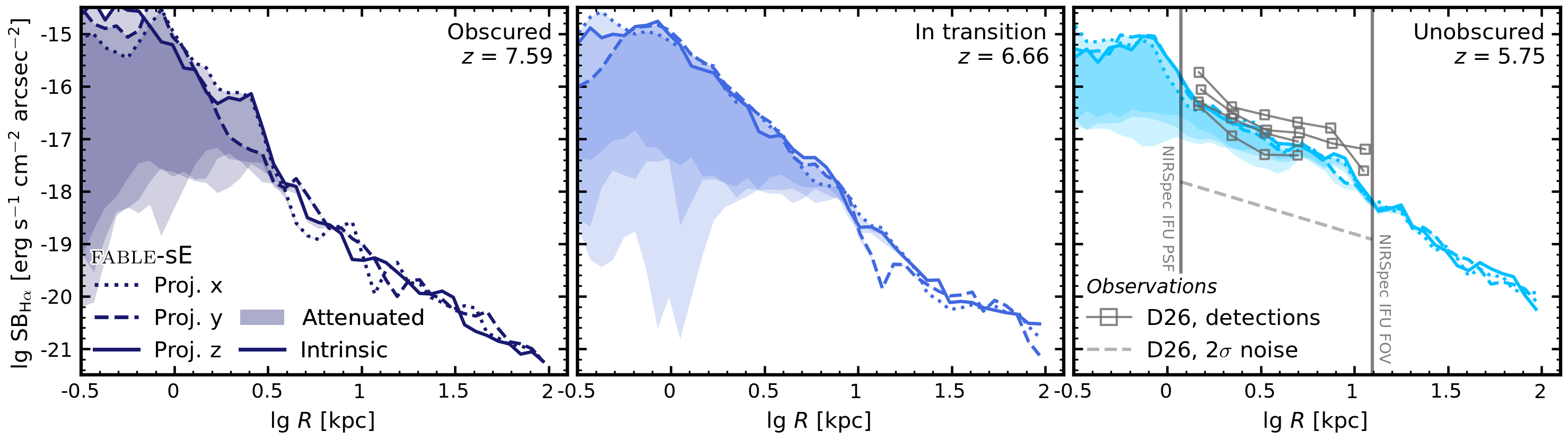}
    \caption{Radial profiles of ${\rm H\alpha}$ surface brightness for the three distinct obscuration stages in the \sE{} run: obscured (\textit{left panel}), in transition (\textit{middle panel}) and unobscured (\textit{right panel}). The different line styles show profiles obtained for three orthogonal lines of sight (arbitrarily chosen along the `x', `y', and `z' directions of the simulation volume). The thick lines display the intrinsic emission, while the shaded areas below indicate the predicted attenuation by a dust screen placed in front of the emission across the three viewing angles. The grey square data points show $\rm H\alpha$ surface brightness profiles from the BEES sample \citep[][D26, see main text for details]{Durovcikova+_2026_BEES}. We include the $2\sigma$ noise level for the NIRSpec IFU with a light-grey dashed line. We also show the radial region inside of which PSF residuals dominate and the edge of the NIRSpec IFU field-of-view (FOV) with dark-grey vertical continuous lines, as labelled on the right panel. \emph{The steepness of the slope of the simulated surface brightness profiles decreases as the quasar becomes unobscured. Attenuation by dust significantly decreases the inner ($R\lesssim3~{\rm kpc}$) surface brightness profiles while leaving the outer parts unchanged, with the strength of attenuation also decreasing as the quasar obscuration decreases. There is also variation along different lines of sight, particularly when the quasar is in a transitional stage, as some viewing angles may be heavily obscured while others are mostly clear. The profiles are generally in agreement with recent observations of $z\sim6$ quasars.}}
    \label{fig:radial_profiles}
\end{figure*}

The ionization state of the CGM is distinct for the different simulations. In NoAGN, the high column densities of gas in the centre ($N_{\mathrm{H}}\gtrsim10^{24}$\,cm$^{-2}$) shield much of the outgoing radiation, leaving most of the neutral gas in the halo untouched. The radiation only escapes through limited lines of sight, and the resulting H$\alpha$ maps are relatively faint, with only some (filamentary) gas illuminated. In the simulations with black holes, the cumulative effect of AGN feedback reduces central column densities and clears out channels through which the ionizing radiation can escape. A much larger portion of the halo and cosmic filaments is ionized, leading to extended emission in H$\alpha$. Focusing on \sE{}, the H$\alpha$ emission is much more diffuse and reveals the disrupted morphology of gas in the halo. The fragmented filaments and gas clumps are illuminated by the central quasar and become partially ionized, which allows them to produce $\rm H \alpha$ emission and contribute to nebular emission. The activity and strength of AGN feedback is thus closely linked to the distribution and extent of $\rm H\alpha$ emission and the response of the CGM.

In Fig.~\ref{fig:f_hi-sE-obs}, we provide a visual overview of the radiative transfer simulations at three distinct obscuration stages for the \sE{} run, previously selected based on the median column density of gas in the central 5~kpc surrounding the black hole (see Fig.~\ref{fig:obscuration_detail}). Here, we use a constant luminosity $\langle Q_{\hi{}} \rangle$ that is set such that the same total energy as in the variable, time-dependent light curve $Q_{\hi{}}(t)$ is injected over the quasar period $t_{\rm quasar}$, i.e.:
\begin{equation}
    \langle Q_{\hi{}} \rangle= \frac{\int_{t_{\rm snap}}^{t_{\rm snap}+t_{\rm quasar}} Q_{\hi{}} (t) \, dt}{t_{\rm quasar}} \, ,
\end{equation}
where $Q_{\hi{}}(t)$ is the time-dependent ionizing rate, $t_{\rm snap}$ corresponds to the cosmic age at the snapshot that is being post-processed (e.g., $t_{\rm snap}\approx841.5~{\rm Myr}$ at $z=6.6$), and $t_{\rm quasar}\approx1~{\rm Myr}$ is the duration of the quasar phase. $Q_{\hi{}}(t)$ is obtained from converting the light curve of the simulations $L_{\rm bol}(t)$ using spectra from the \textsc{qsosed} model (see Appendix~\ref{app:ionizing-photon-rate}).

The constant ionizing rates used here are: $\langle Q_{\hi{}} \rangle_{\rm obscured}=2.89\times10^{57}~{\rm s^ {-1}}$, $\langle Q_{\hi{}} \rangle_{\rm in~ transition}=1.40\times10^{56}~{\rm s^ {-1}}$, and $\langle Q_{\hi{}} \rangle_{\rm unobscured}=1.33\times10^{56}~{\rm s^ {-1}}$. Each row of Fig~\ref{fig:f_hi-sE-obs} shows the neutral hydrogen fraction, neutral hydrogen column density and H$\alpha$ surface brightness for `obscured', `in transition' and `unobscured' phases. 

At $z\approx7.59$ (\textit{upper row of Fig.~\ref{fig:f_hi-sE-obs}}), the quasar is heavily obscured and enshrouded by very dense gas, with a median hydrogen column density $N_{\mathrm{H}}\gtrsim10^{25}$\,cm$^{-2}$, and a significant fraction of the halo is neutral. After an extended bright quasar phase lasting $t_{\rm{quasar}}\approx1~{\rm Myr}$ at $\langle Q_{\hi{}} \rangle \approx3\times10^{57}~{\rm s^{-1}}$, the halo is not strongly affected by the radiation, and we only see H$\alpha$ emission along one direction. This can be explained by the column density varying by several orders of magnitude along different lines of sight, such that some radiation can escape the galaxy, but only along a very narrow solid angle. During that phase, the central engine could not be detected and would have little radiative impact on the surrounding gaseous halo.

During a transitional stage at $z \approx 6.66$ (\textit{middle row of Fig.~\ref{fig:f_hi-sE-obs}}), significant gas accretion flow obscures the central black hole, while AGN feedback is clearing out lower-density channels out of which radiation can escape. In fact, some of the radiation escapes out to the IGM ($R\gtrsim100~{\rm kpc}$) and partially ionizes some of the filaments, which contribute to emission in ${\rm H\alpha}$ that is much more extended than when the central engine is obscured. Nonetheless, the majority of the emission is concentrated in the central region that is immediately irradiated by the quasar. We also note that the propagation of radiation, as traced by the ${\rm H\alpha}$ emission, remains anisotropic, given that feedback from the AGN also operates anisotropically along the directions of least resistance \citep{costa+_2014}. 

The unobscured stage at $z=5.75$ (\textit{lower row of Fig.~\ref{fig:f_hi-sE-obs}}) is characterised by the lowest central hydrogen column densities ($N_{\mathrm{H}}\sim10^{22}$\,cm$^{-2}$) and the smallest scatter across different lines of sight. The continuous and sustained AGN feedback of \sE{} has ejected most of the gas from the centre, so that the radiation can easily escape and photoionize the entire halo. Some structures, e.g. a `blob' in the centre-right, remain shielded from the radiation, while most of the gas is ionized by the quasar. We see extended and clumpy emission in ${\rm H \alpha}$, which reflects the structure of gas in \sE{} following repeated outflow episodes from strong AGN feedback.

In Fig~\ref{fig:radial_profiles}, we show annulus-averaged radial profiles of $\rm H\alpha$ surface brightness for the three obscuration stages of the \sE{} run, obtained from the maps in Fig.~\ref{fig:f_hi-sE-obs}. To assess the impact of the chosen line of sight, we include three orthogonal projections. We also show the predicted profiles assuming that the intrinsic emission is attenuated by a dust screen (see Section~\ref{sec:methods} for details). We include recent observations from the BEES survey \citep[][]{Durovcikova+_2026_BEES} in the rightmost panel. The sample consists of five bright $z\sim6$ quasars ($L_{\rm bol}\approx7.4\times10^{45}$--$1.3\times10^{48}~{\rm erg~s^{-1}}$), four of which have detections of nebular emission and one does not. We display the surface brightness profiles of each detection after converting measured angular offsets to physical kpc, assuming the quasars are at $z=6$. For two of the quasars, the profiles do not extend to the outermost radial bins, attributed to shorter lifetimes, while the two other quasars extend to $R\gtrsim10~{\rm kpc}$; see details in \citet[][]{Durovcikova+_2026_BEES}. We include the $2\sigma$ noise level and FOV for the NIRSpec IFU as labelled on the figure.

We see that the slope of the surface brightness profiles is less steep as the quasar becomes unobscured, consistent with what is inferred from the maps. For the obscured stage (\textit{left panel}), the central surface brightness is high in the centre and rapidly drops off below detection levels, decreasing to ${\rm SB_{\rm H\alpha}}\lesssim10^{-19}~{\rm erg~s^{-1}~cm^{-2}~arcsec^{-2}}$ by $R\approx6~{\rm kpc}$. During the transitional stage (\textit{middle panel}), the profile reaches this value at $R\approx10~{\rm kpc}$. When the quasar is unobscured (\textit{right panel}), the surface brightness profile is bright and remains above this threshold out to $R\approx30~{\rm kpc}$. One caveat of this analysis, however, is that we compare the different obscuration stages at different cosmic times. In particular, the central black hole, the galaxy and the host halo significantly grow between $z=7.59$ and $z=5.75$, and some of the differences in the surface brightness profiles may arise from this evolution. One would have to compare the obscured and unobscured stages at the same time (e.g., $z=5.75$) to more quantitatively disentangle these effects, though we note that this cannot be self-consistently achieved with only one object.

The simulated profiles are generally in agreement with observations from the BEES quasars. The quasars with shorter lifetimes and smaller observed $\rm H\alpha$ surface brightness radial extent almost overlap with the profiles for the unobscured quasar in \sE{}. For the quasars with more extended profiles, the simulations predict slightly lower $\rm H\alpha$ surface brightness, likely because the resolution of the simulations or the ISM modelling start to impact how well smaller substructures that are likely to be illuminated are modelled \citep[e.g.,][]{Bennett+_2026}.

The level of attenuation by dust also depends strongly on the obscuration stage. It is systematically higher when the quasar is obscured and systematically lower when it is unobscured. In the transitional stage, we note that there is a lot of variation from one viewing angle to another, which is consistent with Fig.~\ref{fig:obscuration_detail} wherein some lines of sight are heavily obscured while others are not. Additionally, the attenuation mostly affects the central ($R\lesssim7~{\rm kpc}$) surface brightness while the outer parts are largely unchanged. In fact, the regions where attenuation is strongest are observationally inaccessible due to PSF residuals dominating within $R\leq1$--$2~{\rm kpc}$ at $z=6$ \citep[][]{Durovcikova+_2026_BEES}.

\section{Discussion}\label{sec:discussions}

In this section, we discuss the results obtained in this paper and outline various caveats of the simulations that are relevant to this work.

\subsection{The effect of different SED models for quasar emission}

Recent works suggest that the SED of super-Eddington accreting SMBHs could differ from that of standard quasars, which may explain some peculiar properties of \textit{JWST}-detected high-redshift AGN \citep[such as their intrinsic X-ray weakness, e.g.,][]{madau_haardt_2024, Pacucci_Narayan_2024, inayoshi_kimura_noda_2025, Trinca+_2026}. To tentatively explore this issue, we compare the predicted \hi{} ionizing photon rate for a $10^{10}~{\rm M_{\sun}}$ black hole accreting at $f_{\rm Edd}=1,2,3$ for the \textsc{qsosed} model and for spectra from \citet[][]{Trinca+_2026}. 
We find that the \citet[][]{Trinca+_2026} model produces a similar number of ionizing photons and thus indistinguishable surface brightness maps and profiles compared to the \textsc{qsosed} model. However, the theoretical underpinning of the models is different, and we anticipate that the properties of higher ionization lines and for X-rays will be affected more significantly. Future work comparing more varied SED models, including observational and parametric models \citep[e.g.,][]{Shen+_2020, Temple_Hewett_Banerji_2021}, across more ionization lines should address this and determine whether extended emission during bright quasar phases can help distinguish between different SED choices.

\subsection{Emission line nebulae as probes of high-redshift quasar activity}

High-redshift quasar lifetimes and duty cycles inferred from proximity zones are extremely short ($t_{\rm quasar}\lesssim 10^4$--$10^5~{\rm yr}$), challenging the paradigm wherein accretion feeds the black hole and directly illuminates the surrounding gas, instead requiring strong obscuration and/or perturbation from line of sight effects. There have been alternative attempts to probe quasar lifetimes and past line of sight obscuration at $z\gtrsim5$ by measuring the imprints of the quasar's ionizing radiation in the transverse direction, which should power extended emission-line nebulae \citep[e.g.,][]{Heckman+_1991, Liu+_2013}. \citet[][]{durovcikova+_2025_LyA} search for extended $\rm Ly\alpha$ nebulae around $z\sim6$ quasars and find that the independent measurements agree, though the resonant nature of the $\rm Ly\alpha$ line renders its interpretation more complex \citep[][]{costa+_2022}. More robust measurements come from rest-frame optical emission lines such as $\rm H\alpha$ or [\ion{O}{iii}]$\lambda\lambda4959,5007$, which \textit{JWST} is now targeting \citep[e.g.,][]{Liu+_2025, Durovcikova+_2026_BEES, Wolf+_2026}, confirming the existence of quasars with short lifetimes ($t_{\rm quasar}\lesssim10^5~{\rm yr}$) and providing further empirical support for obscured growth in the early Universe.

The results obtained in this work highlight that the central engine remains undetected and goes through obscured growth for most of its lifetime (see Fig.~\ref{fig:obscuration_detail}). We further show that $\rm H\alpha$ nebular emission directly probes the obscuration stage of the accreting SMBH, with profiles remaining centrally concentrated and dropping off steeply when heavily obscured, even after a sustained bright quasar phase lasting around $1~{\rm Myr}$. While the nebular emission shown in the left panel of Fig.~\ref{fig:radial_profiles} could be observationally attributed to short-lived quasar lifetimes, we rather find that it is associated with a period of intense accretion and radiation output that is heavily obscured. Such constant `lightbulb' scenarios, however, are unlikely to be realistic because quasar light curves show significant variability. Nonetheless, \citet[][]{satyavolu+_2023a_obs} show that quasars with small duty cycle ($<10~{\rm per~cent}$) and short active periods ($t_{\rm quasar}\sim10^4~{\rm yr}$) can reconcile measurements of both small and large proximity zones assuming that much of the black hole's growth is obscured. Future radiative transfer simulations using realistic light curves from the simulations will enable us to study the effect of quasar variability (whose luminosity may vary rapidly by 1--2~dex) on lifetime estimates and provide further theoretical support for obscured growth and short but bright radiative phases in high-redshift quasars.

\subsection{Gas clumps in the CGM of \sE{}}

In this work, we find that large-scale outflows launched by strong AGN feedback from super-Eddington accreting SMBHs significantly perturb and transform the CGM of the quasar host halo at $z\sim6$. We show that radiatively efficient accretion episodes can probe the obscuration stage and/or feedback strength of the central engine, as gas is generally more clumpy in \sE{}, leading to the formation of a brighter and more extended $\rm H\alpha$ nebula.

The clumpiness of the halo is a result of multiple processes that stem from the interaction between AGN-driven outflows and gas in the CGM, including efficient cooling of hot, metal-enriched outflows and shattering of infalling cosmic filaments into several substructures. There remain several exciting open questions concerning the detailed properties of the gas clumps in \sE{}. For instance, the lifetime and eventual fate of these clumps remain uncertain. These systems may also form stars and become `fleeing' sites for star formation outside of the main galaxy, which may be counterparts to the ``cosmic wallflowers'' stellar clusters discussed in \citet[][]{vanDonkelaar+_2026}. We provide a detailed investigation into the properties of the gas clumps in \sE{} in future work (van Donkelaar et al. in prep.).

\subsection{Modelling limitations}

\subsubsection{Physics of the ISM}

One serious limitation of this study, shared with many other similar works, is the modelling of the (unresolved) ISM, which is approximated via an `effective' equation of state \citep[][]{springel_hernquist_2003}. A more realistic treatment of the ISM would account for cold, dense, neutral and star-forming clumps embedded in the ambient warm, ionized medium. This could have a major impact on the growth and feedback of black holes, as they would couple differently with multi-phase gas compared to the current subgrid modelling \citep[e.g.,][]{Wagner_Bicknell_Umemura_2012, Wagner_Bicknell_Umemura_2013}. This would also affect the radiative transfer simulations. Here, we manually set $T_{\rm ISM}=10^4~{\rm K}$, such that star-forming/ISM gas cells are almost entirely neutral and comprise a dense neutral cocoon inside of which the quasar resides. The quasar's radiation would naturally couple differently to a multi-phase ISM, further affecting its propagation out into the CGM and IGM. Given the resolution and modelling assumptions presented here, we do not attempt a more complex approach. We note that recent works have included a model for the multiphase ISM \citep[e.g.,][]{lupi+_2019, Lupi+_2022, Lupi+_2024, Trebitsch+_2021, Sanati+_2025, quadri+_2025}, enabling a more detailed assessment of the impact of AGN feedback and radiation at high redshift which future work should address.

\subsubsection{Dust physics}

The simulations performed in this work do not explicitly track dust; instead, we assume that dust is traced by cool, metal-enriched gas \citep[as in][]{bennett+_2024} where appropriate. This represents another caveat of this work, because dust regulates much of the evolution of the ISM and directly affects observational predictions \citep[see detailed discussions in][]{McKinnon+_2017, Byun+_2025, RodriguezMontero+_2026}. Furthermore, we do not model the direct effects of radiation pressure on dusty gas, which is predicted to be an important channel of AGN feedback capable of launching galactic-scale outflows and regulating star formation already at $z\sim6$ \citep[e.g.,][]{Fabian_1999, Murray_Quataert_Thompson_2005, Ishibashi_Fabian_2015, Thompson+_2015, bieri+_2017, costa+_2018a_driving_shells, costa+_2018b_quenching_sf}. This could affect the interplay between the AGN and surrounding gas, as well as the propagation and properties of large-scale outflows, but would require including a detailed dust model and on-the-fly radiative transfer. 

Additionally, we do not include dust in the radiative transfer simulations in post-processing, which we plan to model in future work. Under the adopted assumptions, however, including further attenuation by dust in the radiative transfer would accentuate and further strengthen our conclusions: in the obscured stage, the gas cocoon would shield the quasar even more and potentially block all of the escaping radiation; in the transitional stage, the anisotropy of the gas distribution caused by AGN feedback would be more important, and the escaping radiation would be more directional; finally, when the quasar is unobscured, there is not enough gas to strongly attenuate the emitted radiation in the centre. To mimic the effects of dust on the observability of the $\rm H\alpha$ emission following a quasar episode, we attenuate the profiles assuming that a dust screen is placed in front of the emission. While the central profiles are significantly attenuated, the outer edges are not affected. Furthermore, most of the attenuation remains observationally inaccessible with, e.g., \textit{JWST} because the central regions are PSF-dominated \citep[][]{Durovcikova+_2026_BEES}. Finally, we caution that our estimates for the attenuation are pessimistic and that realistic profiles are likely not as attenuated.

\subsubsection{Numerical resolution}

Previous work has shown that increased spatial and mass resolution may dramatically alter CGM structure on small scales \citep[e.g.,][]{Hummels+_2019, vandeVoort+_2019, Ramesh_Nelson_2024}. This manifests as an increase in the number of small-scale clouds \citep[e.g.,][]{Ramesh+_2026} and a greater quantity of cool gas condensing out of the CGM \citep[e.g.,][]{vandeVoort+_2019, Bennett_Sijacki_2020, Lucchini+_2026} with higher resolution. In our simulations, the limited resolution significantly affects the formation and survival of dense structures in the CGM. This specifically impacts the number, properties and structure of gas clumps, as we cannot track their fragmentation and collapse into finer sub-kpc-scale substructures, and the impact they have on the observational tracers used in this work \citep[e.g., \hi{} covering fraction and $\rm H\alpha$ nebular emission; see also][van Donkelaar et al. in prep., for further discussions]{Bennett+_2026}.

\subsubsection{Self-consistent modelling of radiative effects}

In this study, we have quantified the radiative output of high-redshift quasars in post-processing, representing an improvement in the modelling of the physics of SMBHs in numerical simulations and offering new avenues for the detectability and observational diagnostics of AGN feedback. However, this approach is not self-consistent because the radiation and hydrodynamics do not directly co-evolve \citep[e.g.,][]{Ledos_2026}. Instead, properly investigating this would require re-running the simulations, including the effects of radiation on-the-fly, which is beyond the scope of the current paper. Another caveat of the present study is the lack of radiation from stars which may contribute to the ionization of the CGM; instead, we have opted to model gas as being in equilibrium with a metagalactic UVB, and leave the addition of stellar radiation in the radiative transfer calculations to future work.

There are multiple recent works from the literature that include modelling of radiative effects on-the-fly for high-redshift AGN in differing ways \citep[][]{lupi+_2019, Kannan+_2022_thesan, Bulichi+_2025_lumina, quadri+_2025, Sanati+_2025, Shen+_2026_lumina, Zier+_2026_lumina}. For instance, while both the \textsc{thesan} \citep[][]{Kannan+_2022_thesan} and \textsc{Lumina} \citep[][]{Bulichi+_2025_lumina, Shen+_2026_lumina, Zier+_2026_lumina} projects employ the same sub-grid ISM model from \citet[][]{springel_hernquist_2003}, also adopted in our work, they vary in the implementation of radiative effects depending on the target of the study. In the former, non-equilibrium cooling below the effective equation of state is allowed as star-forming gas is coupled to the radiation, which results in a departure from the black hole model on which it was calibrated. In contrast, the latter models a transparent ISM in which star-forming gas is decoupled from radiation, adhering more faithfully to the IllustrisTNG \citep[][]{Weinberger+_2017_tng, illustristng_2018} implementation and matching the $z=0$ relations on which it was calibrated. The authors note, however, that the \textsc{Lumina} model should be recalibrated to match the constraints of rare high-redshift quasars \citep[][]{Bulichi+_2025_lumina}.

One of the major limitations hindering simulations with radiative transfer on-the-fly remains the unresolved nature of the ISM, because it underpins the interactions between sources (stellar and AGN) and the escape of radiation on extremely small scales, thereby significantly affecting predictive accuracy \citep[e.g.,][]{Smith+_2022}. Current and future higher-resolution simulations, including more sophisticated modelling of the multi-phase ISM and non-equilibrium coupling between matter and radiation \citep[e.g.,][]{Weinberger_Hernquist_2023, Bollati_Weinberger_2026}, will be essential to advance our physical understanding of these mechanisms. For example, this approach has been adopted in the \textsc{serra} simulations \citep[][]{Pallottini+_2022} and for the re-runs in \textsc{thesan-zoom} \citep[][]{Kannan+_2025_thesan_zoom} to more realistically interpret new ALMA and \textit{JWST} observations, albeit without the inclusion of black hole physics at the moment.

\section{Summary and conclusions}\label{sec:conclusions}

We have analysed a suite of three zoom-in simulations targeting the most massive $z\sim6$ halo from the Millennium cosmological volume \citep[][]{springel+_2005} to investigate the response of the CGM and observational signatures of strong AGN feedback at high redshift. The suite includes a simulation without black hole physics (NoAGN), one with the fiducial \fable{} galaxy formation model, and a modified simulation promoting early and enhanced black hole growth (\sE{}). The quasar host halo, embedded in a large-scale overdensity, lies at the intersection of multiple cosmic filaments that provide a steady supply of cold gas. In \sE{}, provided that the SMBH is seeded early enough and allowed to accrete up to mildly super-Eddington rates, it will power a quasar with a luminosity of up to $L_{\rm bol}\sim 10^{48}~{\rm erg~s^{-1}}$ at early cosmic times. The growth of such extreme objects was previously found to severely impact gas in and out of the halo as a result of enhanced AGN feedback \citep[][]{bennett+_2024}. We quantitatively investigated how AGN feedback shapes the CGM and performed radiative transfer simulations of quasar phases using the novel ray-tracing code \vl{} in post-processing. Here, the main aim was to track ionized hydrogen to estimate how well this and associated optical recombination lines such as H$\alpha$ trace feedback episodes driven by a high-redshift quasar. This is particularly timely given that current and upcoming surveys with \textit{JWST}, such as the \textit{Aether} project (Farina et al., 2026, in prep.), are actively probing the diversity of quasar environments at $z\gtrsim6$. Our main conclusions are:

\begin{itemize}

    \item[$\bullet$] Despite the early onset of quasar activity, the central engine remains heavily obscured throughout most of its lifetime, with many lines of sight being Compton-thick down to $z\lesssim6$. Our work thus suggest that we are likely missing an abundant population of heavily obscured quasars at $z>6$. However, successive episodes of AGN feedback eventually clear out several escape channels in a protracted `transition' stage, ultimately depleting the central reservoir through a strong `blow-out' event lasting $\approx200~{\rm Myr}$, leaving the engine exposed and unobscured. The demographics of unobscured/blue quasars at $z \gtrsim 6$ can hence serve as a powerful constraint on the mechanisms and strength of AGN feedback and how it regulates quasar obscuration.
    
    \item[$\bullet$] During the `blow-out' phase, massive and powerful large-scale outflows induce cosmic filament shattering, and drive dense, cold, outflowing ($v_{\rm rad}>1300~{\rm km~s^{-1}}$) and metal-enriched gas `clumps' out to and beyond the virial radius. This outflow \textit{transforms} the CGM in \sE{}, which temporarily acquires a much higher covering fraction of neutral hydrogen in the CGM (up to $\sim80~{\rm per~cent}$ more than in the NoAGN run). Our results indicate that observations probing the CGM of $z \gtrsim 6$ quasars may play a major role in distinguishing between alternative BH growth pathways in the early Universe.
    
    \item[$\bullet$] We find that quasar phases significantly boost hydrogen ionized fractions in the CGM and IGM up to $\gtrsim100~{\rm kpc}$ scales. The associated H$\alpha$ emission provides a distinctive tracer of both AGN feedback strength and AGN obscuration state. During the `unobscured' phase, the reduced central gas column densities enable more ionizing radiation to escape into the CGM, illuminating the dense, predominantly neutral clumps produced during the `blow out' phase. These photoionized structures subsequently contribute to the formation of brighter and more spatially extended ${\rm H\alpha}$ nebulae that may be detected out to $R\approx30~{\rm kpc}$. 

    \item[$\bullet$] Simulated radial profiles of $\rm H\alpha$ emission agree well with recent observations of $z\sim6$ quasars, suggesting that the observed quasar populations at $z>6$ may have experienced significant feedback. Future \textit{JWST} observations of extended nebular emission in the rest-frame optical will help further constrain quasar radiative activity at high redshift.
\end{itemize}

As observations of high-redshift quasars and their environment continue to probe a greater number of these objects at earlier cosmic epochs \citep[accelerated by Euclid, see e.g.,][]{Belladitta+_2026_Euclid, Yang_D+_2026_Euclid} with ever-increasing depth, it becomes increasingly important for theoretical and simulation work to directly model this wealth of data. This can only be achieved through improved modelling of AGN physics, and, in particular, their radiative output. The simulations presented in this work mark an important step forward in connecting the strength of AGN feedback to observational tracers, revealing the formation of extended emission-line nebulae in $\rm H\alpha$ following sustained bright quasar phases. To further scrutinise black hole physics at high redshift, it will be necessary for future simulations to include a more realistic treatment of the multiphase ISM and its interaction with the quasar radiation field. Explicit modelling of the cold phase will also enable access to complementary observational diagnostics which, together with current and future observations from, e.g., ALMA and \textit{JWST}, will be crucial to constrain the growth and feedback of extremely massive black holes within the first billion years of the Universe.

\section*{Acknowledgements}
We thank the anonymous referee for comments and suggestions that improved the clarity and quality of this manuscript. LT is grateful to Martin Haehnelt, Laura Keating and Chris Done for insightful discussions. LT thanks Matthew Smith and Angus Beane for providing \textsc{vortrace}\footnote{See \url{https://vortrace.readthedocs.io/en/latest/index.html}} to create projection maps of \arepo{} simulations. LT is grateful to Stephen Wilkins for sharing \textsc{qsosed} spectra, and to Alessandro Trinca and Alessandro Lupi for sharing their SED models. LT acknowledges support from the Science and Technology Facilities Council (STFC) for a PhD studentship (STFC Quota Award - ST/Y509139/1). DS acknowledges support from the Science and Technology Facilities Council (STFC) under grant ST/W000997/1. JSB acknowledges support from a Leverhulme Trust Early Career Fellowship. This work was performed through DiRAC DP317 and DP379 projects, using resources provided by: the DiRAC@Durham facility managed by the Institute for Computational Cosmology on behalf of the STFC DiRAC HPC Facility (\url{www.dirac.ac.uk}). The equipment was funded by BEIS capital funding via STFC capital grants ST/P002293/1, ST/R002371/1, and ST/S002502/1, Durham University and STFC operations grant ST/R000832/1. DiRAC is part of the National eInfrastructure; the Cambridge Service for Data Driven Discovery (CSD3) operated by the University of Cambridge Research Computing Service (\url{www.csd3.cam.ac.uk}), provided by Dell EMC and Intel using Tier-2 funding from the Engineering and Physical Sciences Research Council (capital grant EP/P020259/1).

\section*{Data Availability}
 
The data underlying this article will be shared upon reasonable request to the corresponding author.


\bibliographystyle{mnras}
\bibliography{references}



\appendix

\section{The neutral fraction of hydrogen in ionization equilibrium}\label{app:neutral-hydrogen}

\begin{figure}
    \centering
    \includegraphics[width=\linewidth]{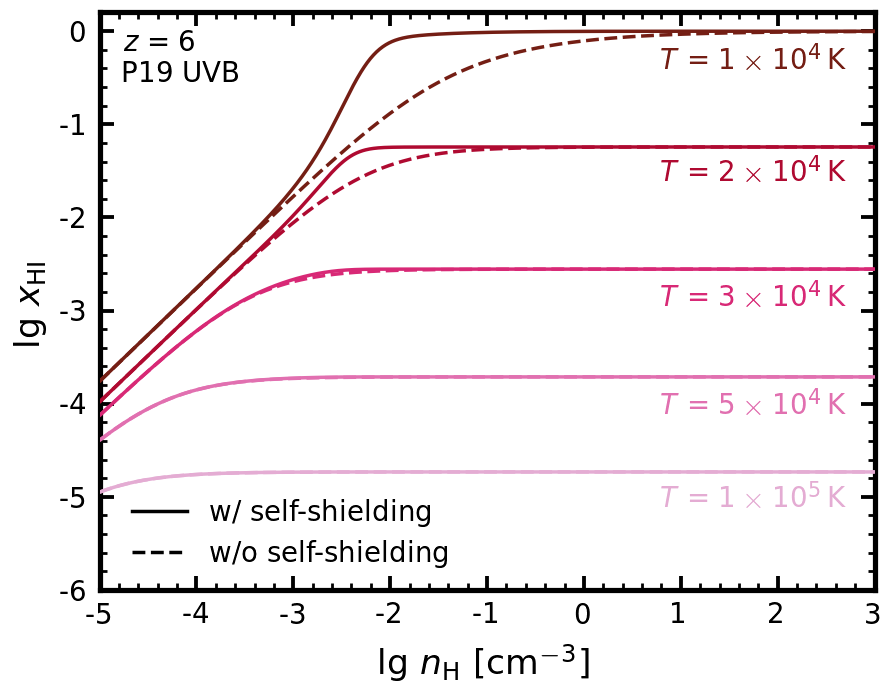}
    \caption{Ionization-equilibrium neutral hydrogen fraction $x_{\hi{}}$ as a function of hydrogen number density $n_{\ion{H}{}}$ for varying temperature at redshift $z=6$. The solid (dashed) lines indicate the neutral fraction obtained with (without) self-shielding as in \citet[][]{rahmati+_2013}. We select $\Gamma_{\mathrm{UVB}}$ from the `equivalent-equilibrium' model of \citealt{puchwein+_2019} (P19). \emph{At low temperatures, the functional form for} $x_{\hi{}}$ \emph{displays a strong dependence on the hydrogen number density; self-shielding enhances the neutral hydrogen fraction with a transition to} $x_{\hi{}}\sim 1$ \emph{for} $n_{\ion{H}{}}\gtrsim n_{\mathrm{self-shield}}$. \emph{At temperatures $\gtrsim2$--$3\times10^4$\,}K\emph{, the neutral hydrogen fraction is very low as a result of collisional ionization.}}
    \label{fig:xHi_nH_T}
\end{figure}

In this Section, we describe how the ionization state of hydrogen is determined for every gas cell in the simulation. We assume gas to be in ionization equilibrium with a meta-galactic UVB, and derive the hydrogen neutral fraction from the internal energy per unit mass $u_{\mathrm{int}}$ (specific internal energy) and density $n$ of the gas extracted from the simulations \citep[in a manner analogous to][see their Appendix A2]{rahmati+_2013}.

We neglect the contribution from helium in this computation, which is a common simplification adopted in the literature \citep[e.g.,][]{f-g+_2009, Altay+_2011, rahmati+_2013} that also suits our approach for evolving the ionization state of hydrogen in post-processing\footnote{We have verified that using a more complex network including helium \citep[e.g.,][]{katz_weinberg_hernquist_1996} does not change our results.}. In equilibrium, the hydrogen ionization fraction is obtained by equating the rate of recombinations with the total ionization rate (see equation~\eqref{eq:hydrogen-ionization-fraction}):
\begin{equation}\label{eq:ioni_eq}
    n_{\hi{}} \left(\Gamma + C_{\hi{}} n_{\rm e} \right) =\alpha_{\ion{H}{ii}} n_{\rm e} n_{\ion{H}{ii}}\, ,
\end{equation}
where $n_{\hi{}}, \, n_{\ion{H}{ii}}$ and $n_{\rm e}$ are the neutral hydrogen, ionized hydrogen and free electron number density, respectively; $\alpha_{\ion{H}{ii}}$ is the recombination rate; $\Gamma$ is the photoionization rate; and $C_{\hi{}}$ is collisional ionization coefficient.

Let us define the hydrogen neutral fraction $x_{\hi{}}\equiv n_{\hi{}}\,/\,n_{\ion{H}{}}$ such that $x_{\ion{H}{ii}}=1-x_{\hi{}}$. Since Helium has been excluded, $n_{\mathrm{e}}=n_{\ion{H}{ii}}$. We can then recast equation~\eqref{eq:ioni_eq} into:
\begin{equation}\label{eq:simpl_ioni_eq}
    x_{\hi{}} \left(\Gamma + C_{\hi{}} n_{\rm e} \right) = \alpha_{\ion{H}{ii}}(1-x_{\hi{}})^2 n_{\ion{H}{}}\,.
\end{equation}
Equation~\eqref{eq:simpl_ioni_eq} can be rearranged to give an exact expression for $x_{\hi{}}$, provided we assume an explicit form for the photoionization, collisional ionization and recombination rates.

In the absence of quasar radiation, photoionization is modelled via a homogeneous, redshift-dependent meta-galactic UVB that may be attenuated due to self-shielding. We use the fitting function provided by \citet[][]{rahmati+_2013}:
\begin{equation}\label{eq:phot_rate}
    \frac{\Gamma}{\Gamma_{\mathrm{UVB}}}=(1-f)\left[1+\left(\frac{n_{\ion{H}{}}}{n_0}\right)^{\beta} \right]^{\alpha_1} + f\left[1 + \frac{n_{\ion{H}{}}}{n_0} \right]^{\alpha_2} \, .
\end{equation}
The various parameters $n_0, \, \alpha_1, \, \alpha_2$ and $\beta$ respectively characterise the density at which self-shielding occurs and the shape of the photoionization curve, and they are fitted as a function of redshift, which was found to reproduce the photoionization rate distribution from radiative transfer simulations \citep[][]{rahmati+_2013, chardin_kulkarni_haehnelt_2018}; and $\Gamma_{\mathrm{UVB}}$ is the photoionization rate due to the UV background at a given redshift. There exists a plurality of UV background models in the literature \citep[e.g.,][]{haardt_madau_1996, haardt_madau_2012, f-g+_2009, khaire_srianand_2019, puchwein+_2019, f-g_2020} with varying redshift evolution calibrated on different observations. In this study, we use the `equivalent-equilibrium' photoionization rates \citep[see][Appendix D2]{puchwein+_2019}, which are intended for use in cosmological hydrodynamical simulations that assume ionization equilibrium. 

The chosen recombination rate is the case B rate $\alpha_{\ion{H}{ii}}\equiv\alpha_B(\lambda(T))$, with $\lambda(T):=315614\,/\,T$, provided by the fitting function found in \citet[][]{Hui_Gnedin_1997}:
\begin{equation}\label{eq:rec_rate}
    \alpha_B(T)=2.753\times10^{-14} \ \frac{\lambda(T)^{1.5}}{[1+(\lambda(T)\,/\,2.74)^{0.407}]^{2.242}} \ \mathrm{cm^3\,s^{-1}} \, .
\end{equation}

Collisional ionization is parametrised through $C_{\ion{H}{i}} \equiv \Lambda (\lambda(T))$, where $\Lambda(T)$ is given in \citet[][]{Hui_Gnedin_1997}:
\begin{equation}\label{eq:coll_rate}
    \Lambda(T) = 5.85 \times 10^{-11} \, \frac{\sqrt{\, T \,} \, \exp(-157809 \, / \, T)}{1+\sqrt{\, T \, / \, 10^5}}~{\rm cm^3\,s^{-1}} \, .
\end{equation}

By combining equations~\eqref{eq:phot_rate},~\eqref{eq:rec_rate},~and~\eqref{eq:coll_rate}, we obtain the following closed-form expression for the equilibrium neutral hydrogen fraction \citep[][]{rahmati+_2013}:
\begin{equation}\label{eq:neutral_hydrogen_frac}
    x_{\hi{}}=\frac{B-\sqrt{B^2-4AC}}{2A} \, ,
\end{equation}
where $A=\alpha_A + \Lambda$, $B=2\alpha_A + \Gamma \,/\, n_{\ion{H}{}} + \Lambda$ and $C=\alpha_A$. This expression is explicitly dependent on the hydrogen number density and temperature of every gas cell (through the recombination and collisional ionization rates, and through the self-shielding from photoionization), and redshift (through its dependence in $\Gamma_{\mathrm{UVB}}$), such that $x_{\hi{}} \equiv x_{\hi{}}(n_{\ion{H}{}}, T, z)$.

There are two additional subtleties in this computation which relate to the temperature dependence of $x_{\hi{}}$. As highlighted in Section~\ref{sec:methods}, we manually set $T_{\rm ISM}=10^4~{\rm K}$ such that $x_{\hi{}}\approx1$ for star-forming gas. There is also a practical issue with the computation: the simulations do not explicitly track the temperature but instead record the internal energy $u_{\mathrm{int}}$ and mean molecular weight $\mu_{\mathrm{sim}}$ of every gas cell, from which the temperature $T_{\mathrm{sim}}$ can be computed. The method introduced above computes the neutral hydrogen fraction using $T_{\mathrm{sim}}$, and the electron abundance will be modified accordingly. As such, we obtain a modified value of the mean molecular weight $\mu_{\mathrm{mod}}$ which can be used to compute a modified temperature $T_{\mathrm{mod}}$. This modified temperature may be different from that used to initially estimate the neutral hydrogen fraction (i.e., $T_{\mathrm{sim}}\neq T_{\mathrm{mod}}$). To avoid this circularity argument, we iteratively solve the equation. Effectively, we start with fully-ionized gas (i.e., $n_{\mathrm{e}}=n_{\ion{H}{}}$) and stop when the values of $n_{\mathrm{e}}$ in two successive iterations differ by less than 10$^{-5}$ per cent \citep[see also][]{katz_weinberg_hernquist_1996}. For every gas cell, the initial fraction of neutral hydrogen is obtained via equation~\eqref{eq:neutral_hydrogen_frac}, whose explicit dependence on the hydrogen number density and temperature at $z=6$ is shown in Fig.~\ref{fig:xHi_nH_T}.

\section{The ionizing photon rate of neutral hydrogen of AGN}\label{app:ionizing-photon-rate}

\begin{figure}
    \centering
    \includegraphics[width=1.0\linewidth]{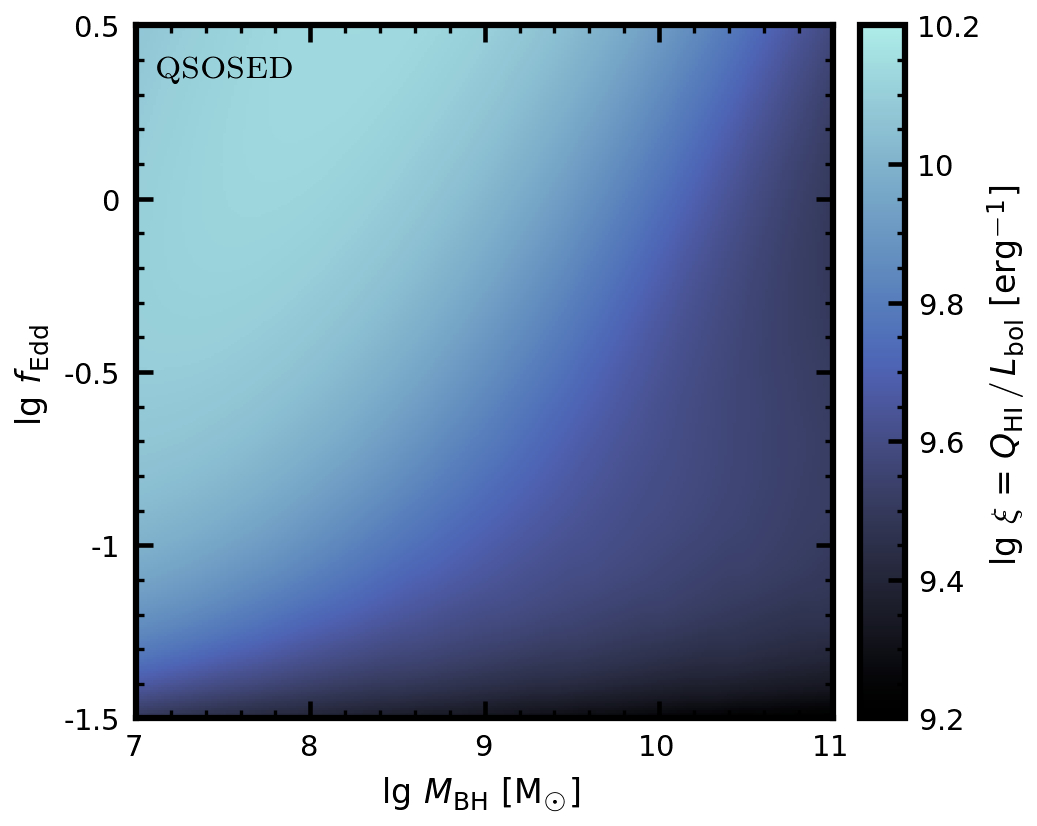}
    \caption{Bolometric correction grid $\xi(\mbh,\fedd)$ for the \textsc{qsosed} model, computed by integrating the SED for a given black hole mass and accretion rate above $E_{\rm ion, \hi{}}=13.6~{\rm eV}$. \emph{We highlight that the model predicts up to $\sim1~{\rm dex}$ variation in the ionizing photon rate production depending on black hole mass and accretion rate.}}
    \label{fig:ionizing-photon-rate-summary}
\end{figure}

In principle, to obtain the ionizing photon rate at any time, we need to produce the SED corresponding to the black hole mass and accretion rate at the time, and integrate it as in equation~\eqref{eq:ioni-photon-rate}. Performing this operation for every time-step of the simulation is computationally expensive. Instead, in this work, we follow an approach similar to \citet[][]{Wilkins+_2025} using the \textsc{synthesizer} framework \citep[][]{Lovell+_2025_Synthesizer, Roper+_2026_Synthesizer}, wherein we pre-compute bolometric correction grids $\xi\equiv\xi(\mbh, \fedd)$ such that:
\begin{equation}\label{eq:ioni-photon-grid}
    Q_{\hi{}}(t)=\xi(\mbh,\fedd) \ L_{\mathrm{bol}}(t) \, .
\end{equation}
Specifically, we compute the ionizing photon rate from the SED of accreting SMBHs for varying black hole mass (51 log-spaced bins for $\mbh \in[10^6-10^{11}]~\msun$) and accretion rate (21 log-spaced bins for $\fedd\in[0.03-3]$),\footnote{This range is suitable for our simulations, given that the central SMBH does not spend a significant portion of time at $\fedd\lesssim0.03$ \citep[][]{bennett+_2024} and, by construction, $\fedd\leq2$.} normalised by bolometric luminosity (i.e., we perform the integral in equation~\eqref{eq:ioni-photon-rate} to obtain $\xi=Q_{\hi{}} \, / \, L_{\rm bol}$). We then obtain a grid of values for $\xi$, which we interpolate to have a smooth bolometric correction from $L_{\rm bol}$ to $Q_{\hi{}}$ as a function of time. The grid for the \textsc{qsosed} model is shown in Fig.~\ref{fig:ionizing-photon-rate-summary}, highlighting that the ionizing photon rate can vary by $\sim 1~{\rm dex}$ depending on the accretion rate and black hole mass (independently of the bolometric luminosity of the quasar).


\bsp
\label{lastpage}
\end{document}